\documentclass[revsymb,aps,prb,10pt]{revtex4}

\begin{document}
\title{Symmetry adapted finite-cluster solver for quantum Heisenberg model in
two-dimensions: a real-space renormalization approach}
\author{V.E. Sinitsyn, I.G. Bostrem, A.S. Ovchinnikov}
\address{Department of Physics, Ural State University, 620083, Ekaterinburg, Russia}
\date{\today }

\begin{abstract}
We present a quantum cluster solver for spin-$S$ Heisenberg model on a
two-dimensional lattice. The formalism is based on the real-space
renormalization procedure and uses the lattice point group-theoretical
analysis and nonabelian $SU(2)$ spin symmetry technique. The exact
diagonalization procedure is used twice at each renormalization group step.
The method is applied to the spin-half antiferromagnet on a square lattice
and a calculation of local observables is demonstrated. A symmetry based
truncation procedure is suggested and verified numerically.
\end{abstract}
\maketitle

\section{Introduction}

Low-dimensional magnetic systems are currently a subject of intensive
experimental and theoretical work. Cluster methods, which approximate the
physics of the infinite system by solving the problem for a corresponding
finite cluster, are the most frequently used theoretical approaches, as they
account short-range correlations on the scale of the cluster size. Numerical
standard methods in the field, such as quantum Monte Carlo (QMC), exact
diagonalization (ED) \cite{Dagotto}, and density matrix renormalization
group (DMRG) \cite{White,Peschel}, are able to give essentially exact
results on limited size systems and form a versatile methodological triad in
simulations of model Hamiltonians.

Even though these techniques have had spectacular successes in calculating
ground state energies and many other properties of one-dimensional (1D) and
two-dimensional (2D) quantum spin systems \cite
{Schollwock,Xiang,Farnell,Sandvik} there is a problem with an utilizing of
symmetries and good quantum numbers of the Hamiltonian, which may be
exploited to thin out Hilbert space by decomposing it into a sum of sectors.
Common symmetries and conservation laws encountered in spin systems are: (i)
Ising or XY symmetry (magnetization conservation $S_{tot}^z$=const); (ii)
point group symmetry (parity, angular momentum conserved); (iii) full $SU(2)$
symmetry ($S_{tot}^2$ conserved). Among these symmetries only the first is
usually exploited in numerical calculations. The full $SU(2)$ spin symmetry
is rather hard to implement, since it requires efforts similar to  the
diagonalization of the actual Hamiltonian to construct the eigenstates of $%
S_{tot}^2$. An implementation of nonabelian $SU(2)$ spin symmetry based on
Clebsch-Gordan transformations and elimination of quantum numbers via the
Wigner-Eckart theorem was performed for the interaction round a face (IRF)
models in the framework of the IRF-DMRG method \cite{Nishino}. This
technique has been successfully applied to the spin-$1/2$ Heisenberg chain
and, later, to the spin 1 and 2 Heisenberg chains \cite{Wada}. The
performant DMRG method conserving a total spin quantum number has been
suggested by McCulloch and Gulasci \cite{McCulloch,Gulasci}. An application
of $SU(2)$ symmetries for the matrix product method (MPM) closely related to
the DMRG \cite{Ostlund,Dukelsky}\ gives a rotationally invariant formulation
valid for spin chains and ladders \cite{Dukelsky,Roman}.

As for the lattice point symmetry, despite its importance in characteristing
energy states of a spin system, there appears to have been little previous
work on the subject. Even though an implementing this symmetry does not lead
to a drastic reduction of a dimension of the Hilbert space sector to be
diagonalized we can resolve properties as a function of additional quantum
numbers (irreducible representations of the point group). This circumstance
might be crucial for efficient truncation of Hilbert space in algorithms
based on real-space renormalization group (RSRG) procedure. This generates a
motivation for the present paper, namely, we present a finite cluster solver
based on RSRG scheme which allows to exploit both the continuous nonabelian $%
SU(2)$ symmetry and discrete symmetry of the lattice point group in
application to isotropic two-dimensional {\it spin-}$S$ systems. As an
example illustrating features of our method we consider the spin-1/2
Heisenberg antiferromagnet (AFH) on a square lattice. This choice is
motivated by two reasons. First, the physical properties of the $S=1/2$ AFH
model on the infinite square lattice at $T=0$ have been much studied and
calculated by various methods by many physicists \cite{Manousakis,Barnes}.
The focus has generally been on the ground state energy and staggered
magnetization although some other quantities have also been computed (see
Ref. \cite{Lin}, for example). Second, the underlying idea of our approach
was first developed by Lin and Campbell in the study of this model system 
\cite{Campbell,Cheng}. Before moving on to the details, we discuss important
aspects of finite-lattice simulations using the ED method regarding the
cluster geometry.

The method of exact diagonalization has been used on the best bipartite
finite square lattices with up to $N=38$ vertices \cite{Masui,Betts}. On
each of a set of finite lattices, the Hamiltonian of the quantum spin model
is diagonalized exactly to find the ground-state energy and the ground state
eigenvector. The ground state properties can then be calculated exactly. The
exact ground-state data for each physical property of the model on all
finite lattices are extrapolated against an appropriate inverse power of $N$
to obtain an estimate of the property on the infinite lattice at zero
temperature \cite{Hasenfratz}. Haan et al. \cite{Haan} showed that certain
parallelogram clusters could produce good results in finite-size exact
diagonalization calculations. Later, Betts et al. developed a grading scheme
of parallelogram tiles of the square lattice that could generate the best
finite clusters \cite{Betts}. From a symmetry point of view, this approach
has an apparent flaw: the point symmetry of a parallelogram cluster does not
match that of the infinte square lattice.

In this respect, a renormalization-group (RG) approach suggested by Lin and
Campbell combining exact diagonalization results with a RG type analysis
seems us to be more promising. The basics idea of their calculations for $2D$
AFH on an $n\times n$ cluster ($n$ is an odd integer) is to divide this
cluster into two parts: an inner $\left( n-2\right) \times \left( n-2\right) 
$ cluster and the perimeter (''outer ring''). The AFH model is firstly
solved for the inner cluster and its ground state is mapped onto a single
effective spin (all the excited states are thrown away). Thus, the problem
is reduced to an effective $1D$ AFH model in which spins on the outer ring
experience antiferromagnetic nearest-neighbor interactions and interact
individually with the effective central spin. The latter plays the role of a
staggered external magnetic field. The procedure is repeated for increasing
values of $n$ and demonstrates convincingly that the staggered magnetic
long-range order exists at zero temperature. We note especially that the
lattice point symmetry holds for all the clusters and their ground state
always has spin $S=1/2$ (not a singlet) according to the Lieb-Mattis theorem 
\cite{Mattis}.

We offer to change real-space RG strategy of the approach making it closer
to DMRG\ methodology. Our treatment begins by dividing a cluster into a
central spin and its environment. In the course of real-space RG iterations
the environment increases (technical details are discussed in the text) and
we determine how coupling between the central spin and the environment
varies. Note especially that we address to the exact diagonalization
procedure twice at each RG step. The first use gives access to states of the
environment and the second one does to those of the whole cluster that
provides its spectrum and observables of interest. Within our RG framework,
local results such as the energy per bond $\varepsilon $ and the staggered
magnetic moment $m$ are measured on the central site.

We have carried out the renormalization procedure through systems of size $%
\sqrt{17}\times \sqrt{17}$, and, in contrast to approach in Ref. \cite
{Campbell}, we keep {\it not only} the ground state. For small clusters ($%
\sqrt{5}\times \sqrt{5}$, $3\times 3$, $\sqrt{13}\times \sqrt{13}$) we use
all of the excited states of the environment found by exact diagonalization,
whereas for the cluster of size $\sqrt{17}\times \sqrt{17}$ we apply a
symmetry based truncation procedure, retaining only the states with largest
weight in the environment density matrix. For this cluster we have compared
the exact diagonalization result for $\varepsilon $ and $m$ with those
obtained via our renormalization group, and we regard the resulting better
than $10^{-2}$ $\%$\ agreement as support for the reliability of our
calculations.

We note that several other methods to improve the RSRG calculations have
been previously formulated to study low-energy properties of spin lattice
models. Among the most important and successful ones, one may cite the Real
Space Renormalization Group with Effective Interactions (RSRG-EI) \cite
{Malrieu}, and the Dressed Cluster Method (DCM) \cite{Wind,Hajj}. The first
method is an improvement of the RSRG method originally proposed by Wilson.
By considering the blocks of lattice it extracts effective interactions
between the blocks through the exact diagonalization of dimers of blocks.
The knowledge of the exact spectrum of the dimers enable one to define
interblock effective interactions via an effective Hamiltonian. This
procedure is iteratevely repeated to blocks of blocks providing at very low
cost reasonable estimate of the energy per bond for 1D and 2D spin lattices.
The second method (DCM) uses a single reference wave function as do the
Coupled Cluster Method (CCM) \cite{Bishop,Hale}. This wave function is used
as a\ bath in which a finite cluster is embedded and treated exactly. The
effect of excitations occuring on the bonds around the cluster is taken into
account through a dressing of the cluster configuration interaction (CI)
matrix. This approach gives results for the cohesive energy of the same
accuracy as the best QMC ones. The DCM can be seen as a convenient
approximation of the CCM. However, the problem is formulated as a
diagonalization of a dressed CI matrix instead of the resolution of a
nonlinear system of equations.

The paper is organized as follows: The general formalism for two-dimensional
spin-$S$ systems is introduced in the following section. In Sec. III\ we
apply the method to the AFH model on a square lattice. Finally, our
conclusions and an outlook are presented in Sec. IV.

\section{Two-dimensional isotropic Heisenberg spin-$s$ system.}

\subsection{Cluster states and observables}

In the first step one must identify the cluster. As detailed above, care
should be taken to ensure that the cluster has the same point-group symmetry
as the lattice. Since calculation of antiferromagnetism requires bipartite
clusters, we select a cluster with a bipartite environment of the central
site (the case of this violation will be illustrated in the example of the
cluster $\sqrt{13}\times \sqrt{13}$).

The cluster Hamiltonian

\begin{equation}
\hat{H}=J\sum\limits_{n\vec{\delta}}\vec{S}_n\vec{S}_{n+\vec{\delta}}=\hat{H}%
_u+\hat{V}  \label{Hful}
\end{equation}
is composed of the term $\hat{V}=J\vec{S}_0\sum\limits_{\vec{\delta}}\vec{S}%
_{0+\vec{\delta}}$ describing interactions of the central spin $\vec{S}_0$
with the nearest neighbors at distances $\vec{\delta}$ and rest terms
denoted as the Hamiltonian of the ''environment'' $\hat{H}_u$. Since, by
construction, the cluster retains a lattice point symmetry, its states $%
\left| iSM\Gamma \mu \right\rangle $ with the energies $E_{iS\Gamma }$ are
labeled by the cluster total spin $S$ with the third component $M$ and by
the irreducible representation $\Gamma \mu $ of the cluster point group.
Different states with the same values $SM$ and $\Gamma \mu $ are
distinguished by the index $i$. In addition we need to consider the operator 
$O_{q1}^{1A_1}=\sum\limits_{\vec{\delta}}\left( S_{0+\vec{\delta}}\right)
_q^1$ as a double irreducible tensor which transforms according to identity
representation $A_1$. The same arguments enable us to use the irreducible
form of the central spin operator $\left( S_0\right) _q^1\equiv \left(
S_0\right) _{q1}^{1A_1}$. The part $\hat{V}$ may be written as the inner
product 
\[
\hat{V}=J\sum\limits_q(-1)^q\left[ 
\begin{array}{ccc}
A_1 & A_1 & A_1 \\ 
1 & 1 & 1
\end{array}
\right] \left( S_0\right) _{q1}^{1A_1}O_{-q1}^{1A_1}\equiv \left[ \left(
S_0\right) ^{1A_1}\times O^{1A_1}\right] _{01}^{0A_1}, 
\]
where $\left[ 
\begin{array}{ccc}
A_1 & A_1 & A_1 \\ 
1 & 1 & 1
\end{array}
\right] =1$ is the Clebsch-Gordan coefficient of the cluster point group 
\cite{Koster}.

Let us suppose that we have found the eigenvalues $E_{i_uS_u\Gamma _u}$ and
the eigenstates of the environment Hamiltonian $\hat{H}_u$ in the form $%
\left| i_uS_uM_u\Gamma _u\mu _u\right\rangle $. The basis functions of the
full cluster are obtained by the addition rule of spin angular momentum

\begin{equation}
\left| i_uS_u\Gamma _u;s;SM\Gamma _u\mu _u\right\rangle =\sum\limits_{\mu
_u,\sigma }\left[ 
\begin{array}{ccc}
S_u & s & S \\ 
M_u & \sigma & M
\end{array}
\right] \left| i_uS_uM_u\Gamma _u\mu _u\right\rangle \left| s\sigma
\right\rangle ,  \label{EigenWF}
\end{equation}
where $[\ldots ]$ is a Clebsch-Gordan coefficient, hereinafter we use that
of given in Ref. \cite{Varshalovich}, and $\left| s\sigma \right\rangle $ is
the wave function of the central spin. Since the state $\left| s\sigma
\right\rangle $ is invariant under all transformations of the point symmetry
group, the cluster basis functions transform like that of the environment
according to the same irreducible representations.

The calculation of matrix elements for the Hamiltonian (\ref{Hful}) with the
help of the Wigner-Ekart's theorem yields (see Appendix A) 
\[
\left\langle i_uS_u\Gamma _u;s;SM\Gamma _u\mu _u\left| \hat{H}\right|
i_u^{\prime }S_u^{\prime }\Gamma _u^{\prime };s;S^{\prime }M^{\prime }\Gamma
_u^{\prime }\mu _u^{\prime }\right\rangle = 
\]
\[
=E_{i_uS_u\Gamma _u}\delta _{i_u,i_u^{\prime }}\delta _{S_u,S_u^{\prime
}}\delta _{\Gamma _u,\Gamma _u^{\prime }}\delta _{\mu _u,\mu _u^{\prime
}}\delta _{S,S^{\prime }}\delta _{M,M^{\prime }}+J(-1)^{S_u^{\prime
}+S+1/2}\left\{ 
\begin{array}{ccc}
S_u & s & S \\ 
s & S_u^{\prime } & 1
\end{array}
\right\} \delta _{S,S^{\prime }}\delta _{M,M^{\prime }} 
\]
\begin{equation}
\times \left\langle s\left\| s\right\| s\right\rangle \left\langle
i_uS_u\Gamma _u\left\| O^{1A_1}\right\| i_u^{\prime }S_u^{\prime }\Gamma
_u^{\prime }\right\rangle \delta _{\Gamma _u,\Gamma _u^{\prime }}\delta
_{\mu _u,\mu _u^{\prime }},  \label{HfulME}
\end{equation}
where $\left\{ ...\right\} $ is a $6j$-symbol. The first reduced matrix
element is $\left\langle s\left\| s\right\| s\right\rangle =\sqrt{%
s(s+1)(2s+1)}$ and the latter may be obtained if the environment eigenstates
are known (see Sec.III). The energy per bond is then calculated as

\begin{equation}
\varepsilon _{iS\Gamma _u}=\frac 1z\left( E_{iS\Gamma
_u}-\sum\limits_{i_uS_u}E_{i_uS_u\Gamma _u}\left| \beta _{i_uS_u\Gamma
_u}^{iS\Gamma _u}\right| ^2\right) =\frac 1z\left( E_{iS\Gamma
_u}-\left\langle E_{iS\Gamma _u}^{(env)}\right\rangle \right) ,  \label{ETL2}
\end{equation}
where $z$ is the number of nearest-neighbors of the central spin. The
eigenfunctions 
\begin{equation}
\left| iSM\Gamma \mu \right\rangle =\sum_{i_uS_u}\beta _{i_uS_u\Gamma
_u}^{iS\Gamma }\left| i_uS_u\Gamma ;s;SM\Gamma \mu \right\rangle ,\;(\Gamma
\mu =\Gamma _u\mu _u)  \label{EigenWF1}
\end{equation}
and the energy levels $E_{iS\Gamma }$ are determined by direct
diagonalization of the cluster Hamiltonian $H$ [Eq.(\ref{HfulME})]. The
values $\varepsilon _{iS\Gamma }$ should be regarded as an approximation of
the energy spectrum in the thermodynamical limit, whereas the energy $%
E_{iS\Gamma }$ divided per bond number is much less appropriate for this.

It is important to note that from Eq.(\ref{HfulME}) it follows that to build
the cluster target state $\left| iSM\Gamma \mu \right\rangle $ we need only
to know the states $\left| i_uS_uM_u\Gamma _u\mu _u\right\rangle $ of the
environment with the quantum numbers $\left| S-s\right| \leq S_u\leq S+s$
and $\Gamma _u\mu _u=\Gamma \mu $.

The most important quantity typically measured in numerical simulations is
the ground-state staggered magnetization $M_c$. The quantum mechanical
observable for $z$-projection of the central spin is given as follows

\begin{equation}
\left\langle iSM\Gamma \mu \left| S_0^z\right| iSM\Gamma \mu \right\rangle
=\left( -1\right) ^{1+S+s}M\sqrt{\frac{2S+1}{S(S+1)}}\left\langle s\left\|
S\right\| s\right\rangle \sum_{i_uS_u}\left( -1\right) ^{S_u}\left| \beta
_{i_uS_u\Gamma _u}^{iS\Gamma }\right| ^2\left\{ 
\begin{array}{ccc}
S & 1 & S \\ 
s & S_u & s
\end{array}
\right\} ,  \label{Sz}
\end{equation}
where the identity (\ref{three3j}) is used. The staggered magnetization $M_c$
is determined as 
\[
M_c^2=\lim_{\left| \vec{R}\right| \rightarrow \infty }3\left| \left\langle
S^z(\vec{R})S^z(0)\right\rangle \right| , 
\]
where factor 3 arises from rotational symmetry in spin space. At long
distances $\left| \left\langle S^z(\vec{R})S^z(0)\right\rangle \right|
\approx \left\langle S^z(0)\right\rangle ^2$ that yields our estimate of the
full root-mean-square staggered magnetization per spin $M_c=\sqrt{%
3\left\langle S_0^z\right\rangle ^2}$.

According to Eq.(\ref{Vint}), spin-correlation function in the states of $%
A_1 $-symmetry, the ground state symmetry as shown below, is determined as 
\[
\left\langle iSMA_1\left| S_0^zS_j^z\right| iSMA_1\right\rangle =\frac 13%
\left\langle iSMA_1\left| \vec{S}_0\vec{S}_j\right| iSMA_1\right\rangle 
\]
\begin{equation}
=\frac 1{3z_f}\sum_{i_uS_u}\sum_{i_u^{\prime }S_u^{\prime }}\beta
_{i_uS_uA_1}^{iSA_1}\beta _{i_u^{\prime }S_u^{\prime
}A_1}^{iSA_1}\left\langle s\left\| S\right\| s\right\rangle \left\langle
i_uS_u\left\| S^{1A}\left( r_j\right) \right\| i_u^{\prime }S_u^{\prime
}\right\rangle \left( -1\right) ^{s+S+S_u^{\prime }}\left\{ 
\begin{array}{ccc}
S_u & s & S \\ 
s & S_u^{\prime } & 1
\end{array}
\right\} ,  \label{spincor}
\end{equation}
where $z_f$ is the lattice coordination number. In this calculation it is
convenient to introduce the double irreducible tensor $S_{q1}^{1A}\left(
r_j\right) =\sum\nolimits_j\left( S_j\right) _q^1$ summing spins at distance 
$r_j$, which transforms according to identity representation $A_1$. One can
see that $O_{q1}^{1A_1}=S_{q1}^{1A}\left( \delta \right) .$

\subsection{Increasing cluster size}

As mentioned above, the lattice point-group symmetry should be conserved
with increasing cluster size. The requirement is put into a practical
computational scheme by the following algorithm: (i) At step $N$ we have the
eigenvalues $E_{i_uS_u\Gamma _u}^{(N)}$ and eigenvectors $\left|
i_uS_um_u\Gamma _u\mu _u\right\rangle _{(N)}$ of the environment. Make a
regular symmetry conserving expansion in the cluster size by adding sites
from the next coordination shell. (ii) Using a scheme of coupling of angular
momenta we build the set $\left| i_IS_Im_I\right\rangle $ of states with
total spin $S_I$ and third component $m_I$ for the part that is being
attached to the environment. The index $i_I$ labels other possible quantum
numbers. (iii) In general case, these functions form a basis of reducible
representation of the cluster point group. Based on the projection operator
technique, one build basic functions $\left| i_IS_Im_I\Gamma _I\mu
_I\right\rangle $ transforming according to irreducible representations $%
\Gamma _I\mu _I$. (iv) Using a scheme of coupling of angular momenta build a
new set $\left| i_IS_Im_Ii_{II}S_{II}m_{II};S_um_u\Gamma _u\mu
_u\right\rangle _{(N+1)}$ of states associated to the extended environment,
where the notation $\left| i_{II}S_{II}m_{II}\Gamma _{II}\mu
_{II}\right\rangle =\left| i_uS_um_u\Gamma _u\mu _u\right\rangle _{(N)}$ is
introduced. An interaction between the $N$-th step environment and the part
added to it can be conveniently written through the irreducible tensors $%
U^{1t_u\gamma }$ and $W^{1t_0\gamma }$ built from spin operators of the
''old'' and ''new'' added parts, respectively, 
\[
V=J\sum_{t_ut_0}\sum_{\gamma \nu }\sum_q\left( -1\right) ^q\left[ 
\begin{array}{ccc}
\gamma & \gamma & A_1 \\ 
\nu & \nu & 1
\end{array}
\right] U_{q\nu }^{1t_u\gamma }W_{-q\nu }^{1t_0\gamma }. 
\]
The indices $t_ut_0$ label different tensors of the same symmetry. The
matrix elements of the extended $(N+1)$-th step environment is 
\[
\left\langle i_IS_Im_Ii_{II}S_{II}m_{II};S_um_u\Gamma _u\mu _u\left|
H_u\right| i_I^{\prime }S_I^{\prime }m_I^{\prime }i_{II}^{\prime
}S_{II}^{\prime }m_{II}^{\prime };S_u^{\prime }m_u^{\prime }\Gamma
_u^{\prime }\mu _u^{\prime }\right\rangle =E_{i_uS_u\Gamma _u}^{(N)}\,\delta
_{S_uS_u^{\prime }}\delta _{m_um_u^{\prime }}\delta _{\Gamma _u\Gamma
_u^{\prime }}\delta _{\mu _u\mu _u^{\prime }} 
\]
\begin{equation}
+J\sum\limits_{t_0t_u\gamma }F\left( \Gamma _I\Gamma _{II}\Gamma ;\Gamma
_I^{\prime }\Gamma _{II}^{\prime }\gamma \right) \left( -1\right)
^{S_I^{\prime }+S_{II}+S_u}\left\{ 
\begin{array}{ccc}
S_I & S_{II} & S_u \\ 
S_{II}^{\prime } & S_I^{\prime } & 1
\end{array}
\right\} \left\langle i_IS_I\Gamma _I\left\| U^{1t_I\gamma }\right\|
i_I^{\prime }S_I^{\prime }\Gamma _I^{\prime }\right\rangle \left\langle
i_{II}S_{II}\Gamma _{II}\left\| W^{1t_{II}\gamma }\right\| i_{II}^{\prime
}S_{II}^{\prime }\Gamma _{II}^{\prime }\right\rangle .  \label{Hu}
\end{equation}
The derivation of Eq.(\ref{Hu}) and the reduced matrix elements of the
operators involved in Eq.(\ref{Hu}) are given in Appendix A and Appendix B,
respectively.

At final step, we diagonalize (\ref{Hu}) and find the eigenvalues $%
E_{i_uS_u\Gamma _u}^{(N+1)}$ and eigenvectors 
\[
\left| i_uS_um_u\Gamma _u\mu _u\right\rangle _{(N+1)}=\sum \alpha
_{i_IS_I\Gamma _Ii_{II}S_{II}\Gamma _{II}}^{i_uS_u\Gamma _u}\left[ 
\begin{array}{ccc}
S_I & S_{II} & S_u \\ 
m_I & m_{II} & m_u
\end{array}
\right] \left[ 
\begin{array}{ccc}
\Gamma _I & \Gamma _{II} & \Gamma _u \\ 
\mu _I & \mu _{II} & \mu _u
\end{array}
\right] \left| i_IS_Im_I\Gamma _I\mu _I\right\rangle \left|
i_{II}S_{II}m_{II}\Gamma _{II}\mu _{II}\right\rangle . 
\]
The iteration is closed by recalculating reduced matrix elements of the
irreducible tensors $W^{1t_{II}\gamma }$ in the basis of the extended
environment (see Appendix B). Note that following the scheme we will in some
cases form intermediate clusters, unsuitable for calculations of local
results, with a non-bipartite environment.

\section{An example: spin-1/2 antiferromagnet on a square lattice.}

The spin-half antiferromagnet on a square lattice represents an optimal
playground to study the strength and limitations of the method. To implement
the algorithm, we need first to build wave functions of the environment
which are predetermined by the lattice point symmetry.

To perform calculations we start with the cluster of minimal size $\sqrt{5}%
\times \sqrt{5}$. The sequence of clusters involved in the calculations are
shown in Fig. 1. Within the smallest cluster, the central spin interacts
with the nearest environment consisting of the spins $S_{\alpha _1},S_{\beta
_1},S_{\gamma _1},S_{\eta _1}$. The spin wave functions of the environment
with the total spin number $S_u$ and the third component $M_u$ may be
written as follows 
\[
\left| \frac 12\frac 12(S_{\alpha _1\beta _1})\frac 12\frac 12(S_{\gamma
_1\eta _1})S_uM_u\right\rangle =\sum_{m_{\alpha _1},m_{\beta _1},m_{\gamma
_1},m_{\eta _1}}\sum_{m_{\alpha _1\beta _1},m_{\gamma _1\eta _1}}\left[ 
\begin{array}{ccc}
1/2 & 1/2 & S_{\alpha _1\beta _1} \\ 
m_{\alpha _1} & m_{\beta _1} & m_{\alpha _1\beta _1}
\end{array}
\right] \left[ 
\begin{array}{ccc}
1/2 & 1/2 & S_{\gamma _1\eta _1} \\ 
m_{\gamma _1} & m_{\eta _1} & m_{\gamma _1\eta _1}
\end{array}
\right] \left[ 
\begin{array}{ccc}
S_{\alpha _1\beta _1} & S_{\gamma _1\eta _1} & S_u \\ 
m_{\alpha _1\beta _1} & m_{\gamma _1\eta _1} & M_u
\end{array}
\right] 
\]
\[
\times \left| 1/2m_{\alpha _1}\right\rangle \left| 1/2m_{\beta
_1}\right\rangle \left| 1/2m_{\gamma _1}\right\rangle \left| 1/2m_{\eta
_1}\right\rangle . 
\]
In such a description, all allowed configurations are comprised by a set $%
\left| 00;00\right\rangle $, $\left| 11;00\right\rangle $, $\left|
01;1M\right\rangle $, $\left| 10;1M\right\rangle $, $\left|
11;1M\right\rangle $, $\left| 11;2M\right\rangle $, where we have dropped
the spin $1/2$ arguments for notation convenience. It is easy to see that
the functions $\left| S_{\alpha _1\beta _1}S_{\gamma _1\eta
_1};S_uM_u\right\rangle $ form (in common case) a basis of reducible
representation of the group $D_4$ (for details see Appendix C) 
\[
\hat{g}\left| S_{\alpha _1\beta _1}S_{\gamma _1\eta _1};S_uM_u\right\rangle
=D_{S_{\alpha _1\beta _1}^{\prime }S_{\gamma _1\eta _1}^{\prime },S_{\alpha
_1\beta _1}S_{\gamma _1\eta _1}}^{(S_u)}(\hat{g})\left| S_{\alpha _1\beta
_1}^{\prime }S_{\gamma _1\eta _1}^{\prime };S_uM_u\right\rangle . 
\]
The matrices $D_{\kappa ,{\kappa }^{\prime }}^{(0)}(\hat{g})$ (the upper
index denotes the spin $S_u$) with the multiindex $\kappa =\left\{ S_{\alpha
_1\beta _1}S_{\gamma _1\eta _1}\right\} $ are readily determined and read

\[
D_{\kappa ,{\kappa }^{\prime }}^{(0)}(E)=D_{\kappa ,{\kappa }^{\prime
}}^{(0)}(C_4^2)=D_{\kappa ,{\kappa }^{\prime }}^{(0)}(\sigma _v^{^{\prime
}})=D_{\kappa ,{\kappa }^{\prime }}^{(0)}(\sigma _v^{^{\prime \prime
}})=\left( 
\begin{array}{cc}
1 & 0 \\ 
0 & 1
\end{array}
\right) , 
\]
\[
D_{\kappa ,{\kappa }^{\prime }}^{(0)}(C_4)=D_{\kappa ,{\kappa }^{\prime
}}^{(0)}(C_4^3)=D_{\kappa ,{\kappa }^{\prime }}^{(0)}(C_2^x)=D_{\kappa ,{%
\kappa }^{\prime }}^{(0)}(C_4^2)=\left( 
\begin{array}{cc}
1/2 & -\sqrt{3}/2 \\ 
-\sqrt{3}/2 & -1/2
\end{array}
\right) . 
\]
The functions $\left| 00;\,00\right\rangle $, $\left| 11;\,00\right\rangle $
form a basis of this two-dimensional representation. Still another
representation of $D_4$ can be generated by means of the functions $\left|
01;\,1M\right\rangle $, $\left| 10;1M\right\rangle $ and $\left|
11;1M\right\rangle $ 
\[
D_{\kappa ,{\kappa }^{\prime }}^{(1)}(E)=\left( 
\begin{array}{ccc}
1 & 0 & 0 \\ 
0 & 1 & 0 \\ 
0 & 0 & 1
\end{array}
\right) ,\qquad D_{\kappa ,{\kappa }^{\prime }}^{(1)}(C_4)=\left( 
\begin{array}{ccc}
-1/2 & -1/2 & 1/\sqrt{2} \\ 
-1/2 & -1/2 & -1/\sqrt{2} \\ 
-1/\sqrt{2} & 1/\sqrt{2} & 0
\end{array}
\right) , 
\]
\[
D_{\kappa ,{\kappa }^{\prime }}^{(1)}(C_4^2)=\left( 
\begin{array}{ccc}
0 & 1 & 0 \\ 
1 & 0 & 0 \\ 
0 & 0 & -1
\end{array}
\right) ,\qquad D_{\kappa ,{\kappa }^{\prime }}^{(1)}(C_4^3)=\left( 
\begin{array}{ccc}
-1/2 & -1/2 & -1/\sqrt{2} \\ 
-1/2 & -1/2 & 1/\sqrt{2} \\ 
1/\sqrt{2} & -1/\sqrt{2} & 0
\end{array}
\right) , 
\]
\[
D_{\kappa ,{\kappa }^{\prime }}^{(1)}(C_2^x)=\left( 
\begin{array}{ccc}
1/2 & 1/2 & 1/\sqrt{2} \\ 
1/2 & 1/2 & -1/\sqrt{2} \\ 
1/\sqrt{2} & -1/\sqrt{2} & 0
\end{array}
\right) ,\qquad D_{\kappa ,{\kappa }^{\prime }}^{(1)}{C_2^y}=\left( 
\begin{array}{ccc}
1/2 & 1/2 & -1/\sqrt{2} \\ 
1/2 & 1/2 & 1/\sqrt{2} \\ 
-1/\sqrt{2} & 1/\sqrt{2} & 0
\end{array}
\right) , 
\]
\[
D_{\kappa ,{\kappa }^{\prime }}^{(1)}(\sigma _v^{^{\prime \prime }})=\left( 
\begin{array}{ccc}
-1 & 0 & 0 \\ 
0 & -1 & 0 \\ 
0 & 0 & 1
\end{array}
\right) ,\qquad D_{\kappa ,{\kappa }^{\prime }}^{(1)}(\sigma _v^{^{\prime
}})=\left( 
\begin{array}{ccc}
0 & -1 & 0 \\ 
-1 & 0 & 0 \\ 
0 & 0 & -1
\end{array}
\right) . 
\]
In a similar way we find the matrices $D_{\kappa ,{\kappa }}^{(2)}(\hat{g})$
in the basis $\left| 11;2M\right\rangle $ 
\[
D_{\kappa ,{\kappa }}^{(2)}(\hat{g})=1\;(\forall \hat{g}\in D_4). 
\]
The representations $D^{(S)}$ are the direct sums of the irreducible
representations $D^{(0)}=D^{(0A_1)}\oplus D^{(0B_2)},$ $D^{(1)}=D^{(1B_1)}%
\oplus D^{(1E)},$ $D^{(2)}=D^{(2A_1)}$ (see Appendix E). The basis functions
of these irreducible representations are given by a similarity
transformation 
\begin{equation}
\left| S_uM_u;\Gamma \mu \right\rangle =\sum\limits_{S_{\alpha _1\beta
_1},S_{\gamma _1\eta _1}}\hat{T}_{S_{\alpha _1\beta _1}S_{\gamma _1\eta
_1};\Gamma \mu }^{(S_u)}\left| S_{\alpha _1\beta _1}S_{\gamma _1\eta
_1};S_uM_u\right\rangle ,  \label{Simila}
\end{equation}
and the matrix $\hat{T}_{S_{\alpha _1\beta _1}S_{\gamma _1\eta _1};\Gamma
\mu }^{(S_u)}=\hat{T}_{S_{\alpha _1\beta _1}S_{\gamma _1\eta _1}S{_u}%
^{\prime };S_u\Gamma _u\mu _u}{\delta }_{S_u,S_u^{\prime }}$ found with the
aid of the projection-operator technique reads (for details see Appendix D)

\[
\begin{array}{ccccccc}
& \left| 00;A_11\right\rangle & \left| 00;B_21\right\rangle & \left|
1M;B_11\right\rangle & \left| 1M;E1\right\rangle & \left| 1M;E2\right\rangle
& \left| 2M;A_11\right\rangle \\ 
\left| 00;00\right\rangle & \frac{\sqrt{3}}2 & \frac 12 & 0 & 0 & 0 & 0 \\ 
\left| 11;00\right\rangle & -\frac 12 & \frac{\sqrt{3}}2 & 0 & 0 & 0 & 0 \\ 
\left| 01;1M\right\rangle & 0 & 0 & \frac 1{\sqrt{2}} & \frac 12 & \frac 12
& 0 \\ 
\left| 10;1M\right\rangle & 0 & 0 & \frac 1{\sqrt{2}} & -\frac 12 & -\frac 12
& 0 \\ 
\left| 11;1M\right\rangle & 0 & 0 & 0 & \frac 1{\sqrt{2}} & -\frac 1{\sqrt{2}%
} & 0 \\ 
\left| 11;2M\right\rangle & 0 & 0 & 0 & 0 & 0 & 1
\end{array}
\]
Given the environment eigenfunctions $\left| S_uM_u;\Gamma _u\mu
_u\right\rangle $ with the eigenvalues $E_{S_u\Gamma _{_{{u}}}}$, the
reduced matrix elements of the double irreducible tensor $O^{1A_1}=S_{\alpha
_1}+S_{\beta _1}+S_{\gamma _1}+S_{\eta _1}$ can be computed
straightforwardly using the Wigner-Eckart theorem and the similarity
transformation (\ref{Simila}) 
\begin{equation}
\left[ 
\begin{array}{ccc}
\gamma & \Gamma ^{\prime } & \Gamma \\ 
\nu & \mu ^{\prime } & \mu
\end{array}
\right] ^{*}\left\langle S\Gamma \left\| O^{1\gamma }\right\| S^{\prime
}\Gamma ^{\prime }\right\rangle
=\sum\limits_{S_{12},S_{34}}\sum\limits_{S_{12}^{\prime },S_{34}^{\prime }}%
\hat{T}_{S_{12}S_{34}S;S\Gamma \mu }^{*}\hat{T}_{S_{12}^{\prime
}S_{34}^{\prime }S^{\prime };S^{\prime }\Gamma ^{\prime }\mu ^{\prime
}}\left\langle S_{12}S_{34};S\left\| O_{\ \;\nu }^{1\gamma }\right\|
S_{12}^{\prime }S_{34}^{\prime };S^{\prime }\right\rangle ,  \label{RedMat1}
\end{equation}
where the indeces $\alpha _1$, $\beta _1$, $\gamma _1$, $\eta _1$ are
correspondingly denoted by the numbers 1-4.

To calculate the reduced matrix element that comes into the right-hand side
of Eq.(\ref{RedMat1}) one has to rewrite $O_{q\nu }^{1\gamma }$ through the
spin operators and employ their expressions for the reduced matrix elements
of the spin operators 
\[
\left\langle S_{12}S_{34};S\left\| S_1\right\| S_{12}^{\prime
}S_{34}^{\prime };S^{\prime }\right\rangle =\left( -1\right)
^{1+S_{12}+S_{34}+S_{12}^{\prime }+S^{\prime }}\left[ S_{12},S_{12}^{\prime
},S,S^{\prime }\right] ^{1/2} 
\]
\begin{equation}
\times \left\{ 
\begin{array}{ccc}
S_{12}^{\prime } & 1 & S_{12} \\ 
1/2 & 1/2 & 1/2
\end{array}
\right\} \left\{ 
\begin{array}{ccc}
S^{\prime } & 1 & S \\ 
S_{12} & S_{34} & S_{12}^{\prime }
\end{array}
\right\} \left\langle 1/2\left\| S\right\| 1/2\right\rangle \delta
_{S_{34},S_{34}^{\prime }},  \label{RMS1}
\end{equation}
\[
\left\langle S_{12}S_{34};S\left\| S_2\right\| S_{12}^{\prime
}S_{34}^{\prime };S^{\prime }\right\rangle =\left( -1\right)
^{1+2S_{12}+S_{34}+S^{\prime }}\left[ S_{12},S_{12}^{\prime },S,S^{\prime
}\right] ^{1/2} 
\]
\begin{equation}
\times \left\{ 
\begin{array}{ccc}
S_{12}^{\prime } & 1 & S_{12} \\ 
1/2 & 1/2 & 1/2
\end{array}
\right\} \left\{ 
\begin{array}{ccc}
S^{\prime } & 1 & S \\ 
S_{12} & S_{34} & S_{12}^{\prime }
\end{array}
\right\} \left\langle 1/2\left\| S\right\| 1/2\right\rangle \delta
_{S_{34},S_{34}^{\prime }},  \label{RMS2}
\end{equation}
\[
\left\langle S_{12}S_{34};S\left\| S_3\right\| S_{12}^{\prime
}S_{34}^{\prime };S^{\prime }\right\rangle =\left( -1\right)
^{1+S_{12}+2S_{34}^{\prime }+S}\left[ S_{34},S_{34}^{\prime },S,S^{\prime
}\right] ^{1/2} 
\]
\begin{equation}
\times \left\{ 
\begin{array}{ccc}
S_{34}^{\prime } & 1 & S_{34} \\ 
1/2 & 1/2 & 1/2
\end{array}
\right\} \left\{ 
\begin{array}{ccc}
S^{\prime } & 1 & S \\ 
S_{34} & S_{12} & S_{34}^{\prime }
\end{array}
\right\} \left\langle 1/2\left\| S\right\| 1/2\right\rangle \delta
_{S_{12},S_{12}^{\prime }},  \label{RMS3}
\end{equation}
\[
\left\langle S_{12}S_{34};S\left\| S_4\right\| S_{12}^{\prime
}S_{34}^{\prime };S^{\prime }\right\rangle =\left( -1\right)
^{1+S_{12}+S_{34}+S_{34}^{\prime }+S}\left[ S_{34},S_{34}^{\prime
},S,S^{\prime }\right] ^{1/2} 
\]
\begin{equation}
\times \left\{ 
\begin{array}{ccc}
S_{34}^{\prime } & 1 & S_{34} \\ 
1/2 & 1/2 & 1/2
\end{array}
\right\} \left\{ 
\begin{array}{ccc}
S^{\prime } & 1 & S \\ 
S_{34} & S_{12} & S_{34}^{\prime }
\end{array}
\right\} \left\langle 1/2\left\| S\right\| 1/2\right\rangle \delta
_{S_{12},S_{12}^{\prime }}.  \label{RMS4}
\end{equation}

Since the operator $O^{1A_1}$ coincides with that of the environment total
spin $\hat{S}_u$, it turns out that the matrix elements $\left\langle
S_u\Gamma _u\left| O^{1A_1}\right| S_u^{\prime }\Gamma _u^{\prime
}\right\rangle $ are diagonal 
\[
\left\langle S_u\Gamma _u\left| O^{1A_1}\right| S_u^{\prime }\Gamma
_u^{\prime }\right\rangle =\sqrt{S_u(S_u+1)(2S_u+1)}\delta _{S_u,S_u^{\prime
}}\delta _{\Gamma _u,\Gamma _u^{\prime }}. 
\]
As a consequence, one may check that this property holds for the Hamiltonian
of the total cluster 
\[
\left\langle S_u\Gamma _u;\frac 12;SM\Gamma _u\mu _u\left| \hat{H}\right|
S_u^{\prime }\Gamma _u^{\prime };\frac 12;S^{\prime }M^{\prime }\Gamma
_u^{\prime }\mu _u^{\prime }\right\rangle = 
\]
\[
J(-1)^{S_u^{\prime }+S+1/2}\left\{ 
\begin{array}{ccc}
S_u & 1/2 & S \\ 
1/2 & S_u^{\prime } & 1
\end{array}
\right\} \sqrt{\frac 32S_u(S_u+1)(2S_u+1)}\delta _{S,S^{\prime }}\delta
_{M,M^{\prime }}\delta _{\Gamma _u,\Gamma _u^{\prime }}\delta _{\mu _u,\mu
_u^{\prime }}\delta _{S_u,S_u^{\prime }} 
\]

\begin{table}[t]
\caption{Energies $E_{S\Gamma}$ and $\epsilon_{S\Gamma}$}
\label{cross}%
\begin{ruledtabular}
\begin{tabular}{ccccccccc}
$S\Gamma$ & $\frac 12 A_1$ & $\frac 12 B_1$ & $\frac 12 B_2$ & $\frac 12 E$
& $\frac 32 A_1$ & $\frac 32 B_1$ & $\frac 32 E$ & $\frac 52 A_1$ \\  
$E_{S\Gamma}$ & 0 & -J & 0 & -J & - $\frac 32 J$ & $\frac 12 J$ & $\frac 12
J $ & J \\   
$\epsilon_{S\Gamma}$ & 0 & - $\frac 14 J$ & 0 & - $\frac 14 J$ & - $\frac 38 J$ & $%
\frac 18 J$ & $\frac 18 J$ & $\frac 14 J$%
\end{tabular}
\end{ruledtabular}
\end{table}

A direct calculation shows that the ground state belongs to the Hilbert
space sector with $S=3/2$ and $\Gamma =A_1$. Hence, only the environment
state with $S\Gamma =1A_1$ is needed to find the ground state energy (see
Table I).

Let us now consider the next step, an expansion of the current environment
block due to the next coordination sphere of radius $\sqrt{2}$. After an
addition of four spins $S_a,S_b,S_c,S_d$, the cluster becomes a square of
size $3\times 3$ with the bipartite environment of the central site (Fig.1).
The basis associated with the added part is

\[
\left| \frac 12\frac 12(S_{ab})\frac 12\frac 12(S_{cd})S_IM_I\right\rangle
=\sum_{m_a,m_b,m_c,m_d}\sum_{m_{ab},m_{cd}}\left[ 
\begin{array}{ccc}
1/2 & 1/2 & S_{ab} \\ 
m_a & m_b & m_{ab}
\end{array}
\right] \left[ 
\begin{array}{ccc}
1/2 & 1/2 & S_{cd} \\ 
m_c & m_d & m_{cd}
\end{array}
\right] \left[ 
\begin{array}{ccc}
S_{ab} & S_{cd} & S_I \\ 
m_{ab} & m_{cd} & M_I
\end{array}
\right] 
\]
\begin{equation}
\times \left| 1/2m_a\right\rangle \left| 1/2m_b\right\rangle \left|
1/2m_c\right\rangle \left| 1/2m_d\right\rangle .  \label{Rad2}
\end{equation}
Repeating the basic steps in the approach we obtain the symmetry adapted
basis $\left| S_IM_I;\Gamma _I\mu _I\right\rangle $. The matrix of
corresponding similarity transformation has the form 
\[
\begin{array}{ccccccc}
& \left| 00;A_11\right\rangle & \left| 00;B_11\right\rangle & \left|
1M;B_21\right\rangle & \left| 1M;E1\right\rangle & \left| 1M;E2\right\rangle
& \left| 2M;A_11\right\rangle \\ 
\left| 00;00\right\rangle & \frac{\sqrt{3}}2 & \frac 12 & 0 & 0 & 0 & 0 \\ 
\left| 11;00\right\rangle & -\frac 12 & \frac{\sqrt{3}}2 & 0 & 0 & 0 & 0 \\ 
\left| 01;1M\right\rangle & 0 & 0 & \frac 1{\sqrt{2}} & 0 & \frac 1{\sqrt{2}}
& 0 \\ 
\left| 10;1M\right\rangle & 0 & 0 & \frac 1{\sqrt{2}} & 0 & -\frac 1{\sqrt{2}%
} & 0 \\ 
\left| 11;1M\right\rangle & 0 & 0 & 0 & 1 & 0 & 0 \\ 
\left| 11;2M\right\rangle & 0 & 0 & 0 & 0 & 0 & 1
\end{array}
\]

The environment Hamiltonian includes only interactions between the first and
second coordination spheres 
\begin{equation}
\hat{H}_u=J\left[ \vec{S}_{\alpha_1}\left( \vec{S}_d+\vec{S}_a\right) +\vec{S%
}_{\beta_1}\left( \vec{S}_a+\vec{S}_b\right) +\vec{S}_{\gamma_1}\left( \vec{S%
}_b+\vec{S}_c\right) +\vec{S}_{\eta_1}\left( \vec{S}_c+\vec{S}_d\right)
\right] .  \label{Hsq}
\end{equation}

We now introduce the cluster irreducible tensors $W_{q\mu }^{1\Gamma }$ and $%
U_{q\mu }^{1\Gamma }$ transforming according to representations $\Gamma \mu $
of the point symmetry group $D_4$ (for details see Appendix D)

\[
U_{q1}^{1A_1}=\frac 1{\sqrt{2}}\left( S_{aq}+S_{bq}+S_{cq}+S_{dq}\right)
,\;U_{q1}^{1B_2}=\frac 1{\sqrt{2}}\left( S_{aq}-S_{bq}+S_{cq}-S_{dq}\right)
, 
\]
\[
U_{q1}^{1E}=\frac 1{\sqrt{2}}\left( S_{aq}+S_{bq}-S_{cq}-S_{dq}\right)
,\;U_{q2}^{1E}=\frac 1{\sqrt{2}}\left( S_{aq}-S_{bq}-S_{cq}+S_{dq}\right) , 
\]
\[
W_{q1}^{1A_1}=\frac 1{\sqrt{2}}\left( S_{\alpha _1q}+S_{\beta _1q}+S_{\gamma
_1q}+S_{\eta _1q}\right) ,\;W_{q1}^{1B_1}=\frac 1{\sqrt{2}}\left( S_{\alpha
_1q}-S_{\beta _1q}+S_{\gamma _1q}-S_{\eta _1q}\right) , 
\]
\begin{equation}
W_{q1}^{1E}=\left( S_{\alpha _1q}-S_{\gamma _1q}\right)
,\;W_{q2}^{1E}=\left( S_{\eta _1q}-S_{\beta _1q}\right)  \label{LinComb}
\end{equation}
and then rewrite Eq.(\ref{Hsq}) as

\begin{equation}
H_u=J\sum_{\gamma \nu }\sum_q(-1)^q\left[ 
\begin{array}{ccc}
\gamma & \gamma & A_1 \\ 
\nu & \nu & 1
\end{array}
\right] U_{q\mu }^{1\gamma }W_{-q\mu }^{1\gamma }=J\sum_\gamma \left[
U^{1\gamma }\times W^{1\gamma }\right] _{01}^{0A_1}.  \label{Hu1}
\end{equation}
The reduced matrix elements of the irreducible operators that appear in Eq.(%
\ref{Hu}) can be obtained exactly from the result (\ref{RedMat1})

\begin{equation}
\left\langle S_I\Gamma _I\left\| U^{1A_1}\right\| S_I^{\prime }\Gamma
_I^{\prime }\right\rangle =\frac 1{\sqrt{2}}\left\langle S_I\left\|
S\right\| S_I\right\rangle \delta _{S_I,S_I^{\prime }}\delta _{\Gamma
_I,\Gamma _I^{\prime }},\;\left\langle S_{II}\Gamma _{II}\left\|
W^{1A_1}\right\| S_{II}^{\prime }\Gamma _{II}^{\prime }\right\rangle =\frac 1%
{\sqrt{2}}\left\langle S_{II}\left\| S\right\| S_{II}\right\rangle \delta
_{S_{II},S_{II}^{\prime }}\delta _{\Gamma _{II},\Gamma _{II}^{\prime }},
\label{StartRME}
\end{equation}
\[
\left\langle S_I\Gamma _I\left\| U^{1E}\right\| S_I^{\prime }\Gamma
_I^{\prime }\right\rangle =\left[ 
\begin{array}{cccccc}
& 0A_1 & 0B_1 & 1B_2 & 1E & 2A_1 \\ 
0A_1 & 0 & 0 & 0 & \sqrt{2} & 0 \\ 
0B_1 & 0 & 0 & 0 & -\sqrt{6} & 0 \\ 
1B_2 & 0 & 0 & 0 & -\sqrt{6} & 0 \\ 
1E & -1 & \sqrt{3} & -\sqrt{3} & 0 & -\sqrt{5} \\ 
2A_1 & 0 & 0 & 0 & \sqrt{10} & 0
\end{array}
\right] , 
\]
\[
\left\langle S_{II}\Gamma _{II}\left\| W^{1E}\right\| S_{II}^{\prime }\Gamma
_{II}^{\prime }\right\rangle =\left[ 
\begin{array}{cccccc}
& 0A_1 & 0B_2 & 1B_1 & 1E & 2A_1 \\ 
0A_1 & 0 & 0 & 0 & \sqrt{2} & 0 \\ 
0B_2 & 0 & 0 & 0 & \sqrt{6} & 0 \\ 
1B_1 & 0 & 0 & 0 & -\sqrt{6} & 0 \\ 
1E & -1 & -\sqrt{3} & -\sqrt{3} & 0 & -\sqrt{5} \\ 
2A_1 & 0 & 0 & 0 & \sqrt{10} & 0
\end{array}
\right] . 
\]
To compute matrix elements of the Hamiltonian $H_u$ we construct the basis 
\begin{equation}
\left| i_IS_I\Gamma _I\;i_{II}S_{II}\Gamma _{II};S_uM_u\Gamma _u\mu
_u\right\rangle =\sum_{m_I,m_{II}}\sum_{\mu _I,\mu _{II}}\left[ 
\begin{array}{ccc}
S_I & S_{II} & S_u \\ 
m_I & m_{II} & M_u
\end{array}
\right] \left[ 
\begin{array}{ccc}
\Gamma _I & \Gamma _{II} & \Gamma _u \\ 
\mu _I & \mu _{II} & \mu _u
\end{array}
\right] \left| i_IS_Im_I\Gamma _I\mu _I\right\rangle \left|
i_{II}S_{II}m_{II}\Gamma _{II}\mu _{II}\right\rangle  \label{BWF1}
\end{equation}
formed from the eigenstates $\left| i_IS_Im_I\Gamma _I\mu _I\right\rangle $
and $\left| i_{II}S_{II}m_{II}\Gamma _{II}\mu _{II}\right\rangle $ of the
''new'' and ''old'' added parts, correspondingly. Then we obtain using Eq.(%
\ref{Hutext}) the expression similar to Eq.(\ref{Hu}) with $E_{i_IS_I\Gamma
_I}=E_{i_{II}S_{II}\Gamma _{II}}=0$. Applying exact diagonalization to the
Hamiltonian $H_u$ one can then find the eigenfunctions 
\[
\left| i_uS_u\Gamma _u\mu _u\right\rangle =\sum \alpha _{i_IS_I\Gamma
_I;i_{II}S_{II}\Gamma _{II}}^{i_uS_u\Gamma _u}\left| i_IS_I\Gamma
_I\;i_{II}S_{II}\Gamma _{II};S_uM_u\Gamma _u\mu _u\right\rangle 
\]
and the energy spectrum $E_{i_uS_u\Gamma _u}$ of the environment. By using
the recursion relation (for details see (\ref{B4}) in Appendix B)

\[
\left\langle i_uS_u\Gamma _u\left\| O^{1A_1}\right\| i_u^{\prime
}S_u^{\prime }\Gamma _u^{\prime }\right\rangle =\delta _{\Gamma _u,\Gamma
_u^{\prime }}\sum\limits_{i_I,S_I,\Gamma }\sum\limits_{i_{II}^{\prime
},S_{II}^{\prime },\Gamma _{II}^{\prime }}\sum\limits_{i_{II},S_{II},\Gamma
_{II}}\alpha _{i_IS_I\Gamma _I;i_{II}S_{II}\Gamma _{II}}^{i_uS_u\Gamma
_u}\alpha _{i_IS_I\Gamma _I;i_{II}^{\prime }S_{II}^{\prime }\Gamma
_{II}^{\prime }}^{i_u^{\prime }S_u^{\prime }\Gamma _u^{\prime }} 
\]
\begin{equation}
\times \left( -1\right) ^{1+S_I+S_{II}^{\prime }+S_u}\left[ S_u,S_u^{\prime
}\right] ^{1/2}\left\{ 
\begin{array}{ccc}
S_u & 1 & S_u^{\prime } \\ 
S_{II}^{\prime } & S_I & S_{II}
\end{array}
\right\} \left\langle i_{II}S_{II}\Gamma _{II}\left\| O^{1A_1}\right\|
i_{II}^{\prime }S_{II}^{\prime }\Gamma _{II}^{\prime }\right\rangle
\label{WA1}
\end{equation}
one finds the reduced matrix elements in the environment basis $\left|
i_uS_uA_1\right\rangle $ that come into the matrix of the total cluster (\ref
{HfulME}).

The formulas (\ref{HfulME},\ref{ETL2},\ref{EigenWF1}) allow us to obtain any
of possible $54$ square cluster states. Our calculation shows that the
ground state belongs to the Hilbert space sector with $S=1/2$ and $\Gamma
=A_1$. Hence, only the environment states with $S\Gamma =0A_1$,$1A_1$ are
needed for the evaluation of the ground state energy. Below we summarize the
results obtained for this particular case.

Using Eq.(\ref{Hu}) and the explicit expressions for the nonzero sums of
Clebsch-Gordan coefficients of the point group $D_4$ (see Eq.(\ref{6G}) in
the Appendix A) 
\[
F(A_1A_1A_1;A_1,A_1,A_1)=F(EEA_1;A_1A_1E)=F(EEA_1;EEA_1)=1,\qquad
F(A_1A_1A_1;EEE)=1/2, 
\]
we obtain 
\[
\hat{H_u}^{(0A_1)}=\left[ 
\begin{array}{ccc}
0 & -\frac 1{\sqrt{3}}J & 0 \\ 
-\frac 1{\sqrt{3}}J & -J & -\sqrt{\frac 53J} \\ 
0 & -\sqrt{\frac 53J} & -3J
\end{array}
\right] 
\]
in the basis of the states $\left| 0A_10A_1;00A_1\right\rangle ,\;\left|
1E1E;00A_1\right\rangle ,\;\left| 2A_12A_1;00A_1\right\rangle $. The
diagonalization of $\hat{H_u}^{(0A_1)}$ yields three states of the $0A_1$
symmetry (see Table II) 
\begin{table}[t]
\caption{Environment states of symmetry $0A_1$}
\label{tableI}%
\begin{ruledtabular}
\begin{tabular}{ccccc}
$i_u$ & $\alpha _{0A_10A_1}^{i_u0A_1}$ & $\alpha
_{1E1E}^{i_u0A_1}$ & $\alpha _{2A_12A_1}^{i_u0A_1}$ & $E_{i_u0A_1}$ \\
\hline
$1$ & $0.071$ & $0.449$ & $0.890$ & $-3.651\,J$ \\
$2$ & $0.569$ & $0.715$ & $-0.406$ & $-0.726\,J$ \\
$3$ & $-0.819$ & $0.535$ & $-0.205$ & $0.377\,J$
\end{tabular}
\end{ruledtabular}
\end{table}
As for the $\hat{H_u}$-operator with $S_u=1$, we have the following matrix
representation, with the same considerations as for the $\hat{H_u}^{(0A_1)}$%
-operator, 
\[
\hat{H_u}^{(1A_1)}=\left[ 
\begin{array}{cccc}
0 & 0 & J & 0 \\ 
0 & 0 & J & 0 \\ 
J & J & -\frac 12J & -\frac{\sqrt{5}}2J \\ 
0 & 0 & -\frac{\sqrt{5}}2J & -\frac 52J
\end{array}
\right] 
\]
in the basis $\left| 0B_11B_1;1MA_1\right\rangle ,\;\left|
1B_20B_2;1MA_1\right\rangle ,\;\left| 1E1E;1MA_1\right\rangle ,\;\left|
2A_12A_1;1MA_1\right\rangle $. The states of $1A_1$ symmetry are listed in
Table III. 
\begin{table}[t]
\caption{Environment states of symmetry $1A_1$}
\label{tableII}%
\begin{ruledtabular}
\begin{tabular}{cccccc}
$i_u$ & $\alpha _{0A_31A_3}^{i_u1A_1}$ & $\alpha
_{1A_40A_4}^{i_u1A_1}$ & $\alpha _{1E1E}^{i_u1A_1}$ & $\alpha
_{2A_12A_1}^{i_u1A_1}$ & $E_{i_u1A_1}$ \\ \hline $1$ & $0.153$ &
$0.153$ & $-0.478$ & $-0.851$ & $-3.128\,J$
 \\
$2$ & $-0.470$ & $-0.470$ & $0.566$ & $-0.487$ & $-1.202\,J$
 \\
$3$ & $-0.505$ & $-0.505$ & $-0.672$ & $0.196$ & $1.330\,J$
 \\
$4$ & $-0.707$ & $0.707$ & $0$ & $0$ & $0$
\end{tabular}
\end{ruledtabular}
\end{table}
By using the recursion relation (\ref{WA1}) with the starting value (\ref
{StartRME}), one finds the reduced matrix elements in the environment basis $%
\left| i_uS_uA_1\right\rangle $. Plugging them into Eq.(\ref{HfulME}) we get
the target states $\left| i\frac 12MA_1\right\rangle $ [see Eq.(\ref
{EigenWF1})] of the cluster and their energies $E_{i\frac 12A_1}$ ($i=1..7$%
). The number of states involved in determining the cluster ground state
equals 7 (see Table IV). 
\begin{table}[t]
\caption{Data on the ground state of the cluster $3\times 3$.}
\label{tableIII}%
\begin{ruledtabular}
\begin{tabular}{cccccccc}
$\beta _{1\,0A_1}^g$ & $\beta _{2\,0A_1}^g$ & $\beta _{3\,0A_1}^g$
& $\beta _{1\,1A_1}^g$ & $\beta _{2\,1A_1}^g$ & $\beta
_{3\,1A_1}^g$ & $\beta _{4\,1A_1}^g$ & $E_g$ \\ \hline $-0.712$ &
$0.044$ & $0.010$ & $-0.695$ & $0.0048$ & $-0.090$ & $0.011$ &
$-4.749$
\end{tabular}
\end{ruledtabular}
\end{table}

Now we list the results for observables. The energy per bond found with the
help of Eq.(\ref{ETL2}) is $\varepsilon _g=-0.3442\;J$. This result may be
compared to those results of QMC \cite{Sandvik} $\varepsilon _g=-0.3347J$,
and DMRG $\varepsilon _g=-0.32679J$\ for lattice of size $20\times 20$ and
for number of DMRG states $150$ \cite{Farnell}. (Extropolation of the DMRG\
results in the infinite-lattice limit yields $\varepsilon _g=-0.3321J$). The
best available DCM \cite{Hajj}, CCM \cite{Zeng}\ and RSRG-EI \cite{Malrieu}
results are $-0.33486J,$ $-0.33308J,$ and $-0.33409J$, respectively. Using (%
\ref{Sz}) we get the ground-state expectation value of the z component of
the central spin $\left\langle S_0^z\right\rangle _0=0.173$ and the
staggered magnetization $M=\sqrt{3\left\langle S_0^z\right\rangle _0^2}=0.299
$. For comparison, the extrapolated QMC\ result for the lattice
magnetization $M=0.3070$. We also provide an estimate of the spin-spin
correlation functions (\ref{spincor}) 
\[
\left\langle S_0^zS_{r=1}^z\right\rangle =-0.115,\;\left\langle S_0^zS_{r=%
\sqrt{2}}^z\right\rangle =0.073.
\]
These estimates should be compared with the known results -0.1116 and
0.0637, correspondingly, \cite{Betts}.

We have made a preliminary calculations by using the small cluster $3\times
3 $ and one can see that an accuracy of the results is still insufficient.
However, we have established the following important features:

(i) The ground state of the system belongs to identity representation $A_1$.

(ii) The lowest-lying environment states of the same point symmetry give a
contribution to the ground state of the system with

the largest weight $|\beta _{1\,0A_1}^g|^2+|\beta _{1\,1A_1}^g|^2\approx
0.989$. One can see the coefficients $\beta ^2$ by nothing that the diagonal
matrix elements of the reduced density matrix in DMRG\ language \cite{White}.

(iii) A comparison of the ground state energy per bond as calculated by $%
E_g/12=-0.396J$ and its infinite-lattice approximation (\ref{ETL2}) within
our approach shows that we produce a better result.

At further step, the procedure is repeated and the environment block grows
by adding the coordination sphere of radius $2$. When the new spins $\vec{S}%
_{\alpha _2}$, $\vec{S}_{\beta _2}$, $\vec{S}_{\gamma _2}$, $\vec{S}_{\eta
_2}$ of the sphere are added, the cluster transforms into the rhombus of
size $\sqrt{13}\times \sqrt{13}$. The cluster has the non-bipartite
environment, hence, it is instructive to study this case to examine the
effect of non-biparticity.

The Hamiltonian of the new environment decomposes as 
\begin{equation}
\hat{H}_u=H_u(0)+J\left( \vec{S}_{\alpha _1}\vec{S}_{\alpha _2}+\vec{S}%
_{\beta _1}\vec{S}_{\beta _2}+\vec{S}_{\gamma _1}\vec{S}_{\gamma _2}+\vec{S}%
_{\eta _1}\vec{S}_{\eta _2}\right) .  \label{Hromb}
\end{equation}
$H_u(0)$ contains all interactions within the ''old'' environment, and the
second term describes all couplings between this part and the added sites.

The irreducible tensors built from the added spins are the same as those of
the first coordination sphere (\ref{LinComb}) 
\[
W_{q1}^{1A_1}=\frac 1{\sqrt{2}}\left( S_{\alpha _2q}+S_{\beta _2q}+S_{\gamma
_2q}+S_{\eta _2q}\right) ,\;W_{q1}^{1B_1}=\frac 1{\sqrt{2}}\left( S_{\alpha
_2q}-S_{\beta _2q}+S_{\gamma _2q}-S_{\eta _2q}\right) , 
\]
\begin{equation}
W_{q1}^{1E}=\left( S_{\alpha _2q}-S_{\gamma _2q}\right)
,\;W_{q2}^{1E}=\left( S_{\eta _2q}-S_{\beta _2q}\right) .
\label{LinCombromb}
\end{equation}
One can then cast the Hamiltonian (\ref{Hromb}) in a more amenable form 
\[
\hat{H}_u=\hat{H}_u(0)+\frac 12J\,\left[ U^{1A_1}\times W^{1A_1}\right]
_{01}^{0A_1}+\frac 12J\,\left[ U^{1B_1}\times W^{1B_1}\right] _{01}^{0A_1}+%
\frac 1{\sqrt{2}}J\,\left[ U^{1E}\times W^{1E}\right] _{01}^{0A_1}, 
\]
where $U^{1\gamma }$ are given by

\[
U_{q1}^{1A_1}=\frac 1{\sqrt{2}}\left( S_{\alpha _1q}+S_{\beta _1q}+S_{\gamma
_1q}+S_{\eta _1q}\right) ,\;U_{q1}^{1B_1}=\frac 1{\sqrt{2}}\left( S_{\alpha
_1q}-S_{\beta _1q}+S_{\gamma _1q}-S_{\eta _1q}\right) , 
\]
\begin{equation}
U_{q1}^{1E}=\left( S_{\alpha _1q}-S_{\gamma _1q}\right)
,\;U_{q2}^{1E}=\left( S_{\eta _1q}-S_{\beta _1q}\right) .  \label{LinComb2}
\end{equation}

The matrices formed from the reduced matrix elements of $W^{1\gamma }$
tensor coincide with (\ref{StartRME}). To find those of $U^{1\gamma }$
tensor we use Eq.(\ref{B4}). The expressions mentioned (\ref{StartRME}) are
used to initialize the calculations.

From direct calculations one can show that the quantum numbers $S=5/2$ and $%
\Gamma =A_1$ are attached to the ground state of the rhombus. This state is
formed from $41$ envronment states with the symmetry $S\Gamma _u=2A_1$ and $%
22$ states of symmetry $S\Gamma _u=3A_1$. Numerical diagonalization gives
the cluster ground state energy $E_g(\frac 52A_1)=-5.779J$ that yields the
ground-state energy per bond $\varepsilon _g=-0.30925\;J$ in the
thermodynamic limit. If we compare this result with that of QMC, we see that
the agreemnet becomes worse. Nevertheless, the conclusions made for the
square cluster $3\times 3$ hold: (i) both the ground state of the
environment and that of the total cluster have the lattice point symmetry $%
A_1$. (ii) The largest weight (is of the order $0.993$) into the sum of
diagonal elements in the density matrix comes from three lowest-lying $2A_1$
states and one state of symmetry $3A_1$, whereas the total number of states
is $63$.

Monitoring energies per bond $\varepsilon _{iS\Gamma }$ for the total
cluster spectrum $E_{iS\Gamma }$, we found that the minimal value $%
\varepsilon _{\min }\approx -0.3229\,J$ is reached for the lowest state of
symmetry $\frac 32A_1$, however, $E(\frac 32A_1)>E_g=E(\frac 52A_1)$. A
similar situation, when a minimal energy per bond belongs to a higher lying
state, has been early observed in DMRG study of antiferromagnetic chains 
\cite{White}. Despite the number of sites in the cluster $\sqrt{13}\times 
\sqrt{13}$ is greater than that of in the cluster $3\times 3$, we see that
the result for $\varepsilon _{\min }$ deteriorates compared to the QMC\
value $-0.3347J$. Close inspection allows us to suggest that this is because
we are working on the cluster with a non-bipartite environment.

To proceed with increasing cluster size and satisfy the biparticity
requirement we should take the square cluster $5\times 5$ in the next step.
For the 24-site environment of the cluster, an exact-diagonalization
calculation of the {\it total} spectrum is not possible at present and so,
to move on to the next-larger system, we have to elaborate a procedure for
determining the states giving the best approximation to true environment
states. To solve the problem and implement the condition of bipartite
environment we take a system in the form of ''decorated cross'' obtained
from the former cluster $\sqrt{13}\times \sqrt{13}$ by adding four spins $%
\vec{S}_{\alpha _3}$, $\vec{S}_{\beta _3}$, $\vec{S}_{\gamma _3}$, $\vec{S}%
_{\eta _3}$ (Fig. 1). The form makes equal a number of sites in both
sublattices, though it incorporates 8 sites that are being attached to the
cluster by single lattice bonds. At the same time, exact diagonalization of
the cluster$\sqrt{17}\times \sqrt{17}$ is allowed, hence we compare the
exact diagonalization results with those obtained from a symmetry based
truncation procedure and ananlyze a truncation error on a number of states
kept. Since the cluster increasing is similar to that used in the previous
step, we present only the results of calculations. The ground state of the
extended cluster environment has the symmetry $0A_1$. The total number of
states with the same symmetry is 194. Together with 439 $1A_1$-states of the
environment they form a ground state of the total cluster labeled by the
symmetry numbers $\frac 12A_1$. Results for the ground state energy per bond 
$\varepsilon =-0.3304$, the staggered magnetization $m=0.305$ and the
spin-spin correlation functions $\left\langle S_0^zS_{r=1}^z\right\rangle
=-0.1101$, $\left\langle S_0^zS_{r=\sqrt{2}}^z\right\rangle =0.0615$ agree
well with the mentioned ED and QMC results and are much better than those
obtained for the square cluster $3\times 3$. A deviation from the ED result
is found for $\left\langle S_0^zS_{r=2}^z\right\rangle =0.0169$. This
discrepancy arises from finite size effects and an imperfect topology of the
cluster.

We now describe the low-energy spectrum of the environment. As the dynamics
of N\'{e}el order parameter is the one of a free rotator, the low-energy
levels scale as $E(S)\sim S(S+1)/N$, where the inertia of that rotator is
proportional to the number of sites \cite{Hasenfratz,Misguich}. The
environment lowest-energy levels (tower of states) belonging to different
irreducible representations of the lattice point group are shown in Fig. 2
for different $S$ sectors. The $SU(2)$ breaking due to long-range N\'{e}el
order appears as a set of $A_1$-states, lying off from other levels, with an
energy scaling as $E(S)\sim S(S+1)$.

In the remainder of this section we describe a version of the truncation
procedure. The main idea will be illustrated on an example of the ground
states properties. An inspection of results for the current and previous
clusters reveals that one have to take the lowest-lying environment
eigenstates both in the $0A_1$ and $1A_1$ sectors. As for the number of kept
states it seems to be most simple to take $M$ states equally from the both
subspaces, albeit the choice may not be optimal. To prove that this concept
works we recalculate the observables found above on various number of
envronment states kept (see Table V). As can be seen from Fig. 3 the
convergence of the results is exponentially fast in $M$. Merely keeping 100
basis states may be as efficient as keeping of all 633 environment states
intact. We regard the resulting better than 0.01\% agreement for $%
\varepsilon $ and $\ m$ as support for the efficiency of our truncation
procedure.

\begin{table}[t]
\caption{Convergence of the ground state properties vs number of environment
states kept.}
\label{truncation}%
\begin{ruledtabular}
\begin{tabular}{ccccccc}
$M(0A_1)$ & $M(1A_1)$ & $E_0/J$ & $\varepsilon /J$ & $m$ & $ \left\langle
S_0^zS^z(1)\right\rangle $  & $ \left\langle S_0^zS^z(\sqrt{2})\right\rangle $  \\ \hline
1 & 1 & -7.9010 & -0.2410 & 0.354897 & -0.080333 & 0.065647 \\ 
5 & 5 & -8.1018 & -0.3136 & 0.304148 & -0.104533 & 0.071420 \\ 
10 & 10 & -8.1282 & -0.3238 & 0.304928 & -0.107933 & 0.073201 \\ 
20 & 20 & -8.1378 & -0.3279 & 0.305707 & -0.109300 & 0.073872 \\ 
50 & 50 & -8.1425 & -0.3301 & 0.305101 & -0.110033 & 0.074247 \\ 
100 & 100 & -8.1429 & -0.3303 & 0.305187 & -0.110100 & 0.074289 \\ 
194 & 194 & -8.1430 & -0.3304 & 0.305187 & -0.110133 & 0.074300
\end{tabular}
\end{ruledtabular}
\end{table}

\section{Conclusions.}

In this paper we present a quantum cluster solver for spin-$S$ Heisenberg
model on a two-dimensional lattice. The formalism is based on the real-space
renormalization procedure and uses the lattice point group-theoretical
analysis and nonabelian $SU(2)$ spin symmetry technique. Let us summarize
advantages of the approach:

(i) The cluster spin states are decomposed into parts belonging to different
irreducible representations of the lattice point group and to different
values of the total spin. Due to the embedded group-theoretical analysis,
our approach can handle each of the cluster target states independently that
offers a distinct advantage for parallel computation.

(ii) An extension of MPM destined for quantum spin chains to higher
dimensions has inspired construction of variational methods for the ground
states of 2D spin Hamiltonians (vertex state models \cite{Zittartz}, tensor
product variational approach \cite{Nishino1}, tensor product ansatz \cite
{Martin}). Since, the trial states are represented by two-dimensional
product of local weights, these approaches are face with severe limitations
concerning their applicability because of relation between a spin value and
lattice topology. The shortcoming lacks in our formalism.

(iii) Large sparce-matrix diagonalization algorithms (Lanscoz technique, for
example) used in DMRG and ED methods converge to maximum and minimum
eigenvalues of a model Hamiltonian, i.e. to eigenvalues at the edges of the
spectrum. Our approach gives access to eigenstates of an entire spectrum.

(iv) Combined with decimation procedure of the environment states like those
used in DMRG\ technique the group-theoretical analysis allows us to overcome
exponential growth of computational efforts with increase of system size.
Our approach using the total spin $S$ and the irrep index $\Gamma \mu $ as
good quantum numbers yields a rather reliable truncation procedure of the
Hilbert space of the model Hamiltonian.

(v) Calculation of observables for the central spin involving a density
matrix of the environment reduces edge effects which are inevitable on
finite-size clusters.

The major drawback of the formalism is that it does not allow an easy
implementation: a complexity in construction of basic sets via repeated
evaluation of $6j$ and $6\Gamma $ symbols, the calculation involves two
matrix diagonalizations etc. The performance gains from implementing the $%
SU(2)$ and lattice point symmetries are not impressive in comparison with
gains from exploiting just the simple $U(1)$ symmetry leading to total
magnetization as good quantum number. Their using in studies with larger
clusters without truncation cannot help to alleviate the problem of
exponential growth of computational efforts.

In the method that we suggest, short range correlations on the scale of the
cluster are taken into account, while correlations on a scale larger than
the cluster size are neglected. To overcome this shortcoming we need to
restore translational symmetry of the lattice. The results of these
investigations will be reported elsewhere. In this connection, we note that
the translational invariance holds for the DCM\ and CCM\ methods.

\acknowledgments
We would like to thanks V.V. Valkov and S.G. Ovchinnikov for the useful
discussions. This work was partly supported by the grant NREC-005 of USA
CRDF (Civilian Research \& Development Foundation). One of us (V.E.S.)
thanks the Foundation ''Dynasty'' (Moscow) for the support.

\section{Appendix A}

Let $\left| i_IS_I\Gamma _I\;i_{II}S_{II}\Gamma _{II};SM\Gamma \mu
\right\rangle $ is a state with total spin $S$, third component $M$, and
transforming according to irreducible representation $\Gamma \mu $. This
state appears in the tensor product decomposition $\left( i_IS_I\Gamma
_I\right) \times \left( i_{II}S_{II}\Gamma _{II}\right) $, where $\left(
iS\Gamma \right) $ denotes a state with total spin $S$, irreducible
representation $\Gamma $ and $i$ labels other possible quantum numbers.

We need to compute the matrix element

\[
\left\langle i_IS_I\Gamma _I\;i_{II}S_{II}\Gamma _{II};SM\Gamma \mu \left|
\left[ U^{1\gamma }\times W^{1\gamma }\right] _{01}^{0A_1}\right|
i_I^{\prime }S_I^{\prime }\Gamma _I^{\prime }\;i_{II}^{\prime
}S_{II}^{\prime }\Gamma _{II}^{\prime };S^{\prime }M^{\prime }\Gamma
^{\prime }\mu ^{\prime }\right\rangle 
\]
\[
=\sum_{q\nu }\sum\limits_{\left\{ m,\mu \right\} }\left( -1\right) ^q\left[ 
\begin{array}{ccc}
\gamma & \gamma & A_1 \\ 
\nu & \nu & 1
\end{array}
\right] \left[ 
\begin{array}{ccc}
\Gamma _I & \Gamma _{II} & \Gamma \\ 
\mu _I & \mu _{II} & \mu
\end{array}
\right] ^{*}\left[ 
\begin{array}{ccc}
\Gamma _I^{\prime } & \Gamma _{II}^{\prime } & \Gamma ^{\prime } \\ 
\mu _I^{\prime } & \mu _{II}^{\prime } & \mu ^{\prime }
\end{array}
\right] \left[ 
\begin{array}{ccc}
S_I & S_{II} & S \\ 
m_I & m_{II} & M
\end{array}
\right] \left[ 
\begin{array}{ccc}
S_I^{\prime } & S_{II}^{\prime } & S^{\prime } \\ 
m_I^{\prime } & m_{II}^{\prime } & M^{\prime }
\end{array}
\right] 
\]
\begin{equation}
\times \left\langle i_IS_Im_I\Gamma _I\mu _I\left| U_{q\nu }^{1\gamma
}\right| i_I^{\prime }S_I^{\prime }m_I^{\prime }\Gamma _I^{\prime }\mu
_I^{\prime }\right\rangle \left\langle i_{II}S_{II}m_{II}\Gamma _{II}\mu
_{II}\left| W_{-q\nu }^{1\gamma }\right| i_{II}^{\prime }S_{II}^{\prime
}m_{II}^{\prime }\Gamma _{II}^{\prime }\mu _{II}^{\prime }\right\rangle .
\label{Vint}
\end{equation}
The Wigner-Eckart theorem for a double irreducible tensor reads 
\begin{equation}
\left\langle i_IS_Im_I\Gamma _I\mu _I\left| W_{q\nu }^{1\gamma }\right|
i_I^{\prime }S_I^{\prime }m_I^{\prime }\Gamma _I^{\prime }\mu _I^{\prime
}\right\rangle =(-1)^{S_I-m_I}\left( 
\begin{array}{ccc}
S_I & 1 & S_I^{\prime } \\ 
-m_I & q & m_I^{\prime }
\end{array}
\right) \left[ 
\begin{array}{ccc}
\gamma & \Gamma _I^{\prime } & \Gamma _I \\ 
\nu & \mu _I^{\prime } & \mu _I
\end{array}
\right] ^{*}\left\langle i_IS_I\Gamma _I\left\| W^{1\gamma }\right\|
i_I^{\prime }S_I^{\prime }\Gamma _I^{\prime }\right\rangle ,  \label{WEPG}
\end{equation}
where the $3j$ symbol is related to the Clebsch-Gordan coefficient by 
\[
\left( 
\begin{array}{ccc}
S_1 & S_2 & S_3 \\ 
m_1 & m_2 & m_3
\end{array}
\right) =\frac 1{\sqrt{2S_3+1}}(-1)^{S_I+S_2-m_3}\left[ 
\begin{array}{ccc}
S_1 & S_2 & S_3 \\ 
m_1 & m_2 & -m_3
\end{array}
\right] . 
\]
A full contraction of five Clebsch-Gordan coefficients of the point group
may be written via the $6\Gamma $ symbol

\[
F(\Gamma _I\Gamma _{II}\Gamma ;\Gamma _I^{\prime }\Gamma _{II}^{\prime
}\gamma )=\sum\limits_{\nu \mu _I\mu _{II}\mu _I^{\prime }\mu _{II}^{\prime
}}\left[ 
\begin{array}{ccc}
\gamma & \gamma & A_1 \\ 
\nu & \nu & 1
\end{array}
\right] \left[ 
\begin{array}{ccc}
\Gamma _I & \Gamma _{II} & \Gamma \\ 
\mu _I & \mu _{II} & \mu
\end{array}
\right] ^{*}\left[ 
\begin{array}{ccc}
\Gamma _I^{\prime } & \Gamma _{II}^{\prime } & \Gamma ^{\prime } \\ 
\mu _I^{\prime } & \mu _{II}^{\prime } & \mu ^{\prime }
\end{array}
\right] \left[ 
\begin{array}{ccc}
\gamma & \Gamma _I^{\prime } & \Gamma _I \\ 
\nu & \mu _I^{\prime } & \mu _I
\end{array}
\right] ^{*}\left[ 
\begin{array}{ccc}
\gamma & \Gamma _{II}^{\prime } & \Gamma _{II} \\ 
\nu & \mu _{II}^{\prime } & \mu _{II}
\end{array}
\right] ^{*} 
\]
\begin{equation}
\sim \left\{ 
\begin{array}{ccc}
\Gamma _I & \Gamma _{II} & \Gamma \\ 
\Gamma _{II}^{\prime } & \Gamma _I^{\prime } & \gamma
\end{array}
\right\} \left[ 
\begin{array}{ccc}
A_1 & \Gamma ^{\prime } & \Gamma \\ 
1 & \mu ^{\prime } & \mu
\end{array}
\right] \delta _{\Gamma \Gamma ^{^{\prime }}}\delta _{\mu \mu ^{^{\prime }}},
\label{6G}
\end{equation}
however, it is more convenient to find directly this sum.

Substituting (\ref{WEPG}) into (\ref{Vint}) and performing the sum with the
aid of Eq.(\ref{6G}) and the formula (see \cite{Varshalovich}, for example) 
\[
\sum\limits_{\chi \psi \rho \sigma \tau }\left( -1\right) ^{p-\psi +q-\chi
+r-\rho +s-\sigma +t-\tau }\left( 
\begin{array}{ccc}
p & a & q \\ 
\psi & -\alpha & \chi
\end{array}
\right) \left( 
\begin{array}{ccc}
q & r & t \\ 
-\chi & \rho & \tau
\end{array}
\right) 
\]
\begin{equation}
\times \left( 
\begin{array}{ccc}
r & a & s \\ 
-\rho & \alpha ^{\prime } & \sigma
\end{array}
\right) \left( 
\begin{array}{ccc}
s & p & t \\ 
-\sigma & -\psi & -\tau
\end{array}
\right) =\frac{(-1)^{a-\alpha }}{(2a+1)}\left\{ 
\begin{array}{ccc}
q & p & a \\ 
s & r & t
\end{array}
\right\} \delta _{aa^{\prime }}\delta _{\alpha \alpha ^{\prime }}
\label{4KG}
\end{equation}
we get finally 
\[
\left\langle i_IS_I\Gamma _I\;i_{II}S_{II}\Gamma _{II};SM\Gamma \mu \left|
\left[ U^{1\gamma }\times W^{1\gamma }\right] _{01}^{0A_1}\right|
i_I^{\prime }S_I^{\prime }\Gamma _I^{\prime }\;i_{II}^{\prime
}S_{II}^{\prime }\Gamma _{II}^{\prime };S^{\prime }M^{\prime }\Gamma
^{\prime }\mu ^{\prime }\right\rangle 
\]
\[
=\delta _{SS^{\prime }}\delta _{MM^{\prime }}\delta _{\Gamma \Gamma ^{\prime
}}\delta _{\mu \mu ^{\prime }}\left( -1\right) ^{S_I^{\prime
}+S_{II}+S}\left\{ 
\begin{array}{ccc}
S_I & S_{II} & S \\ 
S_{II}^{\prime } & S_I^{\prime } & 1
\end{array}
\right\} F(\Gamma _I\Gamma _{II}\Gamma ;\Gamma _I^{\prime }\Gamma
_{II}^{\prime }\gamma ) 
\]
\begin{equation}
\times \left\langle i_IS_I\Gamma _I\left\| U^{1\gamma }\right\| i_I^{\prime
}S_I^{\prime }\Gamma _I^{\prime }\right\rangle \left\langle
i_{II}S_{II}\Gamma _{II}\left\| W^{1\gamma }\right\| i_{II}^{\prime
}S_{II}^{\prime }\Gamma _{II}^{\prime }\right\rangle .  \label{Hutext}
\end{equation}
The reduced matrix elements appearing in (\ref{Hutext}) result from the
previous iteration.

\section{Appendix B}

The systematic increasing cluster size requires an iterative procedure to
compute the reduced matrix elements of the double irreducible tensors $%
U^{1\gamma }$ or $W^{1\gamma }$ (acting on the states with indices $I$ and $%
II$, respectively) in the basis 
\begin{equation}
\left| iSm\Gamma \mu \right\rangle =\sum \alpha _{i_IS_I\Gamma
_I;\,i_{II}S_{II}\Gamma _{II}}^{iS\Gamma }\left[ 
\begin{array}{ccc}
S_I & S_{II} & S \\ 
m_I & m_{II} & m
\end{array}
\right] \left[ 
\begin{array}{ccc}
\Gamma _I & \Gamma _{II} & \Gamma \\ 
\mu _I & \mu _{II} & \mu
\end{array}
\right] \left| i_IS_Im_I\Gamma _I\mu _I\right\rangle \left|
i_{II}S_{II}m_{II}\Gamma _{II}\mu _{II}\right\rangle .  \label{B1}
\end{equation}
with aid of the Wigner-Eckart theorem. On the other hand one can use the
basis of states (\ref{B1}) to obtain

\[
\left\langle iSm\Gamma \mu \left| W_{q\nu }^{1\gamma }\right| i^{\prime
}S^{\prime }m^{\prime }\Gamma ^{\prime }\mu ^{\prime }\right\rangle 
\]
\[
=\sum \alpha _{i_IS_I\Gamma _I;\,i_{II}S_{II}\Gamma _{II}}^{iS\Gamma }\alpha
_{i_IS_I\Gamma _I;\,i_{II}^{\prime }S_{II}^{\prime }\Gamma _{II}^{\prime
}}^{i^{\prime }S^{\prime }\Gamma ^{\prime
}}\sum\limits_{m_Im_{II}m_{II}^{^{\prime }}}\left[ 
\begin{array}{ccc}
S_I & S_{II} & S \\ 
m_I & m_{II} & m
\end{array}
\right] \left[ 
\begin{array}{ccc}
S_I & S_{II}^{\prime } & S^{\prime } \\ 
m_I & m_{II}^{\prime } & m^{\prime }
\end{array}
\right] \left( -1\right) ^{S_{II}-m_{II}}\left( 
\begin{array}{ccc}
S_{II} & 1 & S_{II}^{\prime } \\ 
-m_{II} & q & m_{II}^{\prime }
\end{array}
\right) 
\]
\begin{equation}
\times \sum\limits_{\mu _I\mu _{II}\mu _{II}^{^{\prime }}}\left[ 
\begin{array}{ccc}
\gamma _I & \Gamma _{II} & \Gamma \\ 
\mu _I & \mu _{II} & \mu
\end{array}
\right] ^{*}\left[ 
\begin{array}{ccc}
\Gamma _I & \Gamma _{II}^{\prime } & \Gamma \\ 
\mu _I & \mu _{II}^{\prime } & \mu
\end{array}
\right] \left[ 
\begin{array}{ccc}
\gamma & \Gamma _{II}^{\prime } & \Gamma _{II} \\ 
\mu & \mu _{II}^{\prime } & \mu _{II}
\end{array}
\right] ^{*}\left\langle i_{II}S_{II}\Gamma _{II}\left\| W^{1\gamma
}\right\| i_{II}^{\prime }S_{II}^{\prime }\Gamma _{II}^{\prime }\right\rangle
\label{B3}
\end{equation}
The sum over $m_I$, $m_{II}$ and $m_{II}^{^{\prime }}$ is performed with the
aid of the formula 
\[
\sum\limits_{\chi \psi \rho }\left( -1\right) ^{p-\psi +q-\chi +r-\rho
}\left( 
\begin{array}{ccc}
p & a & q \\ 
\psi & \alpha & -\chi
\end{array}
\right) \left( 
\begin{array}{ccc}
q & b & r \\ 
\chi & \beta & -\rho
\end{array}
\right) \left( 
\begin{array}{ccc}
r & c & p \\ 
\rho & \gamma & -\psi
\end{array}
\right) 
\]
\begin{equation}
=\left( 
\begin{array}{ccc}
a & b & c \\ 
-\alpha & -\beta & -\gamma
\end{array}
\right) \left\{ 
\begin{array}{ccc}
a & b & c \\ 
r & p & q
\end{array}
\right\}  \label{three3j}
\end{equation}
The sum of three Clebsch-Gordan coefficients of the lattice point group in
turn can be transformed as follows 
\[
\sum\limits_{\mu _I\mu _{II}\mu _{II}^{^{\prime }}}\sum\limits_{\bar{\nu}%
\bar{\mu}^{\prime }}\left[ 
\begin{array}{ccc}
\Gamma _I & \Gamma _{II} & \Gamma \\ 
\mu _I & \mu _{II} & \mu
\end{array}
\right] ^{*}\left[ 
\begin{array}{ccc}
\Gamma _I & \Gamma _{II}^{\prime } & \Gamma ^{\prime } \\ 
\mu _I & \mu _{II}^{\prime } & \bar{\mu}^{\prime }
\end{array}
\right] \left[ 
\begin{array}{ccc}
\gamma & \Gamma _{II}^{\prime } & \Gamma _{II} \\ 
\bar{\nu} & \mu _{II}^{\prime } & \mu _{II}
\end{array}
\right] ^{*}\delta _{\nu \bar{\nu}}\delta _{\mu ^{\prime }\bar{\mu}^{\prime
}} 
\]
\[
=\sum\limits_{\mu _I\mu _{II}\mu _{II}^{^{\prime }}}\sum\limits_{\bar{\nu}%
\bar{\mu}^{\prime }}\left[ 
\begin{array}{ccc}
\Gamma _I & \Gamma _{II} & \Gamma \\ 
\mu _I & \mu _{II} & \mu
\end{array}
\right] ^{*}\left[ 
\begin{array}{ccc}
\Gamma _I & \Gamma _{II}^{\prime } & \Gamma ^{\prime } \\ 
\mu _I & \mu _{II}^{\prime } & \bar{\mu}^{\prime }
\end{array}
\right] \left[ 
\begin{array}{ccc}
\gamma & \Gamma _{II}^{\prime } & \Gamma _{II} \\ 
\bar{\nu} & \mu _{II}^{\prime } & \mu _{II}
\end{array}
\right] ^{*}\sum\limits_{\bar{\Gamma}\bar{\mu}}\left[ 
\begin{array}{ccc}
\gamma & \Gamma ^{\prime } & \bar{\Gamma} \\ 
\nu & \mu ^{\prime } & \bar{\mu}
\end{array}
\right] ^{*}\left[ 
\begin{array}{ccc}
\gamma & \Gamma ^{\prime } & \bar{\Gamma} \\ 
\bar{\nu} & \bar{\mu}^{\prime } & \bar{\mu}
\end{array}
\right] 
\]
\[
=\sum\limits_{\bar{\Gamma}\bar{\mu}}\left[ 
\begin{array}{ccc}
\gamma & \Gamma ^{\prime } & \bar{\Gamma} \\ 
\nu & \mu ^{\prime } & \bar{\mu}
\end{array}
\right] ^{*}\sum\limits_{\mu _I\mu _{II}\mu _{II}^{^{\prime }}}\sum\limits_{%
\bar{\nu}\bar{\mu}^{\prime }}\left[ 
\begin{array}{ccc}
\Gamma _I & \Gamma _{II} & \Gamma \\ 
\mu _I & \mu _{II} & \mu
\end{array}
\right] ^{*}\left[ 
\begin{array}{ccc}
\Gamma _I & \Gamma _{II}^{\prime } & \Gamma ^{\prime } \\ 
\mu _I & \mu _{II}^{\prime } & \bar{\mu}^{\prime }
\end{array}
\right] \left[ 
\begin{array}{ccc}
\gamma & \Gamma _{II}^{\prime } & \Gamma _{II} \\ 
\bar{\nu} & \mu _{II}^{\prime } & \mu _{II}
\end{array}
\right] ^{*}\left[ 
\begin{array}{ccc}
\gamma & \Gamma ^{\prime } & \bar{\Gamma} \\ 
\bar{\nu} & \bar{\mu}^{\prime } & \bar{\mu}
\end{array}
\right] 
\]

After permutation of the first and second columns in the third
Clebsch-Gordan coefficient the sum over projections $\mu _I$, $\mu _{II}$, $%
\mu _{II}^{^{\prime }}$, $\bar{\nu}$, and $\bar{\mu}^{\prime }$ is easily
performed that gives immediately $6\Gamma $ symbol \cite{Griffith} 
\begin{equation}
\sum\limits_{\bar{\Gamma}\bar{\mu}}\left[ 
\begin{array}{ccc}
\gamma & \Gamma ^{\prime } & \bar{\Gamma} \\ 
\bar{\nu} & \bar{\mu}^{\prime } & \bar{\mu}
\end{array}
\right] ^{*}\left\{ 
\begin{array}{ccc}
\gamma & \Gamma _{II}^{\prime } & \Gamma _{II} \\ 
\Gamma _I & \Gamma & \Gamma ^{\prime }
\end{array}
\right\} \delta _{\Gamma \bar{\Gamma}}\delta _{\mu \bar{\mu}}\,\varepsilon
\left( \Gamma _I\Gamma _{II}^{\prime }\Gamma ^{\prime }\right) ,
\label{sixG}
\end{equation}
where we use the symmetry property of the Clebsch-Gordan coefficients 
\[
\left[ 
\begin{array}{ccc}
\Gamma _1 & \Gamma _2 & \Gamma \\ 
\mu _1 & \mu _2 & \mu
\end{array}
\right] =\varepsilon \left( \Gamma _1\Gamma _2\Gamma \right) \left[ 
\begin{array}{ccc}
\Gamma _2 & \Gamma _1 & \Gamma \\ 
\mu _2 & \mu _1 & \mu
\end{array}
\right] , 
\]
and the sign $\varepsilon \left( \Gamma _1\Gamma _2\Gamma \right) =\pm 1$
depends on the point group.

The reduced matrix element can be computed using (\ref{B3},\ref{sixG}) that
yields the results 
\[
\left\langle iS\Gamma \left\| W^{1\gamma }\right\| i^{\prime }S^{\prime
}\Gamma ^{\prime }\right\rangle =\sum \alpha _{i_IS_I\Gamma
_I;\,i_{II}S_{II}\Gamma _{II}}^{iS\Gamma }\alpha _{i_IS_I\Gamma
_I;\,i_{II}^{\prime }S_{II}^{\prime }\Gamma _{II}^{\prime }}^{i^{\prime
}S^{\prime }\Gamma ^{\prime }} 
\]
\begin{equation}
\times \left( -1\right) ^{1+S_I+S_{II}^{\prime }+S}\left[ S,S^{\prime
}\right] ^{1/2}\left\{ 
\begin{array}{ccc}
S & 1 & S^{\prime } \\ 
S_{II}^{\prime } & S_I & S_{II}
\end{array}
\right\} \left\langle i_{II}S_{II}\Gamma _{II}\left\| W^{1\gamma }\right\|
i_{II}^{\prime }S_{II}^{\prime }\Gamma _{II}^{\prime }\right\rangle \left\{ 
\begin{array}{ccc}
\gamma & \Gamma _{II}^{\prime } & \Gamma _{II} \\ 
\Gamma _I & \Gamma & \Gamma ^{\prime }
\end{array}
\right\} \varepsilon \left( \Gamma _I\Gamma _{II}^{\prime }\Gamma ^{\prime
}\right) ,  \label{B4}
\end{equation}
and 
\[
\left\langle iS\Gamma \left\| U^{1\gamma }\right\| i^{\prime }S^{\prime
}\Gamma ^{\prime }\right\rangle =\sum \alpha _{i_IS_I\Gamma
_I;\,i_{II}S_{II}\Gamma _{II}}^{iS\Gamma }\alpha _{i_I^{\prime }S_I^{\prime
}\Gamma _I^{\prime };\,i_{II}S_{II}\Gamma _{II}}^{i^{\prime }S^{\prime
}\Gamma ^{\prime }} 
\]
\begin{equation}
\times \left( -1\right) ^{1+S_I+S_{II}+S^{\prime }}\left[ S,S^{\prime
}\right] ^{1/2}\left\{ 
\begin{array}{ccc}
S & 1 & S^{\prime } \\ 
S_I^{\prime } & S_{II} & S_I
\end{array}
\right\} \left\langle i_IS_I\Gamma _I\left\| U^{1\gamma }\right\|
i_I^{\prime }S_I^{\prime }\Gamma _I^{\prime }\right\rangle \left\{ 
\begin{array}{ccc}
\gamma & \Gamma _I^{\prime } & \Gamma _I \\ 
\Gamma _{II} & \Gamma & \Gamma ^{\prime }
\end{array}
\right\} ,  \label{B5}
\end{equation}
where $\left[ S\right] \equiv \left( 2S+1\right) $.

\section{Appendix C}

Here, we give a detailed derivation of representation for the $D_4$ point
group in the basis of four coupled connector spins

\[
\left| S_1S_2(S_{12})S_3S_4(S_{34})SM\right\rangle =\sum \left[ 
\begin{array}{ccc}
S_1 & S_2 & S_{12} \\ 
m_1 & m_2 & m_{12}
\end{array}
\right] \left[ 
\begin{array}{ccc}
S_3 & S_4 & S_{34} \\ 
m_3 & m_4 & m_{34}
\end{array}
\right] \left[ 
\begin{array}{ccc}
S_{12} & S_{34} & S \\ 
m_{12} & m_{34} & M
\end{array}
\right] \left| S_1m_1\right\rangle \left| S_2m_2\right\rangle \left|
S_3m_3\right\rangle \left| S_4m_4\right\rangle . 
\]

Consider first the $\pi /2$ rotation $\hat{C}_4$ about the $z$ axis 
\[
\hat{C}_4\left| S_1S_2(S_{12})S_3S_4(S_{34})SM\right\rangle
=\sum_{m_1,m_2}\sum_{m_3,m_4}\sum\limits_{m_{12},m_{34}}\left[ 
\begin{array}{ccc}
S_1 & S_2 & S_{12} \\ 
m_1 & m_2 & m_{12}
\end{array}
\right] \left[ 
\begin{array}{ccc}
S_3 & S_4 & S_{34} \\ 
m_3 & m_4 & m_{34}
\end{array}
\right] \left[ 
\begin{array}{ccc}
S_{12} & S_{34} & S \\ 
m_{12} & m_{34} & M
\end{array}
\right] \left| S_2m_2\right\rangle \left| S_3m_3\right\rangle \left|
S_4m_4\right\rangle \left| S_1m_1\right\rangle 
\]
\[
=\sum_{m_1,m_2}\sum_{m_3,m_4}\sum\limits_{m_{12},m_{34}}\left[ 
\begin{array}{ccc}
S_1 & S_2 & S_{12} \\ 
m_1 & m_2 & m_{12}
\end{array}
\right] \left[ 
\begin{array}{ccc}
S_3 & S_4 & S_{34} \\ 
m_3 & m_4 & m_{34}
\end{array}
\right] \left[ 
\begin{array}{ccc}
S_{12} & S_{34} & S \\ 
m_{12} & m_{34} & M
\end{array}
\right] 
\]
\[
\times \sum_{\bar{m}_1,\bar{m}_2}\sum\limits_{\bar{m}_3,\bar{m}_4}\left| S_2%
\bar{m}_2\right\rangle \left| S_3\bar{m}_3\right\rangle \left| S_4\bar{m}%
_4\right\rangle \left| S_1\bar{m}_1\right\rangle \delta _{\bar{m}%
_2m_2}\delta _{\bar{m}_3m_3}\delta _{\bar{m}_4m_4}\delta _{\bar{m}_1m_1} 
\]
To proceed we use the orthogonality property for the Clebsch-Gordan
coefficients 
\[
\delta _{\bar{m}_2m_2}\delta _{\bar{m}_3m_3}\delta _{\bar{m}_4m_4}\delta _{%
\bar{m}_1m_1}=\sum\limits_{S_{14},m_{14}}\sum\limits_{S_{23},m_{23}}\left[ 
\begin{array}{ccc}
S_2 & S_3 & S_{23} \\ 
m_2 & m_3 & m_{23}
\end{array}
\right] \left[ 
\begin{array}{ccc}
S_2 & S_3 & S_{23} \\ 
\bar{m}_2 & \bar{m}_3 & m_{23}
\end{array}
\right] \left[ 
\begin{array}{ccc}
S_4 & S_1 & S_{14} \\ 
m_4 & m_1 & m_{14}
\end{array}
\right] \left[ 
\begin{array}{ccc}
S_4 & S_1 & S_{14} \\ 
\bar{m}_4 & \bar{m}_1 & m_{14}
\end{array}
\right] . 
\]
Then

\[
\hat{C}_4\left| S_1S_2(S_{12})S_3S_4(S_{34})SM\right\rangle
=\sum_{m_1,m_2}\sum_{m_3,m_4}\sum\limits_{m_{12},m_{34}}\left[ 
\begin{array}{ccc}
S_1 & S_2 & S_{12} \\ 
m_1 & m_2 & m_{12}
\end{array}
\right] \left[ 
\begin{array}{ccc}
S_3 & S_4 & S_{34} \\ 
m_3 & m_4 & m_{34}
\end{array}
\right] \left[ 
\begin{array}{ccc}
S_{12} & S_{34} & S \\ 
m_{12} & m_{34} & M
\end{array}
\right] 
\]
\[
\times \sum_{\bar{m}_1,\bar{m}_2}\sum\limits_{\bar{m}_3,\bar{m}_4}\left| S_2%
\bar{m}_2\right\rangle \left| S_3\bar{m}_3\right\rangle \left| S_4\bar{m}%
_4\right\rangle \left| S_1\bar{m}_1\right\rangle
\sum\limits_{S_{14},m_{14}}\sum\limits_{S_{23},m_{23}}\left[ 
\begin{array}{ccc}
S_2 & S_3 & S_{23} \\ 
m_2 & m_3 & m_{23}
\end{array}
\right] \left[ 
\begin{array}{ccc}
S_2 & S_3 & S_{23} \\ 
\bar{m}_2 & \bar{m}_3 & m_{23}
\end{array}
\right] \left[ 
\begin{array}{ccc}
S_4 & S_1 & S_{14} \\ 
m_4 & m_1 & m_{14}
\end{array}
\right] \left[ 
\begin{array}{ccc}
S_4 & S_1 & S_{14} \\ 
\bar{m}_4 & \bar{m}_1 & m_{14}
\end{array}
\right] 
\]
\[
=\sum_{m_1,m_2}\sum_{m_3,m_4}\sum\limits_{m_{12},m_{34}}\left[ 
\begin{array}{ccc}
S_1 & S_2 & S_{12} \\ 
m_1 & m_2 & m_{12}
\end{array}
\right] \left[ 
\begin{array}{ccc}
S_3 & S_4 & S_{34} \\ 
m_3 & m_4 & m_{34}
\end{array}
\right] \left[ 
\begin{array}{ccc}
S_{12} & S_{34} & S \\ 
m_{12} & m_{34} & M
\end{array}
\right] \sum_{\bar{m}_1,\bar{m}_2}\sum\limits_{\bar{m}_3,\bar{m}_4}\left| S_2%
\bar{m}_2\right\rangle \left| S_3\bar{m}_3\right\rangle \left| S_4\bar{m}%
_4\right\rangle \left| S_1\bar{m}_1\right\rangle 
\]
\[
\times \sum\limits_{S_{14},m_{14}}\sum\limits_{S_{23},m_{23}}\sum\limits_{%
\bar{m}_{14},\bar{m}_{23}}\left[ 
\begin{array}{ccc}
S_2 & S_3 & S_{23} \\ 
m_2 & m_3 & m_{23}
\end{array}
\right] \left[ 
\begin{array}{ccc}
S_2 & S_3 & S_{23} \\ 
\bar{m}_2 & \bar{m}_3 & \bar{m}_{23}
\end{array}
\right] \left[ 
\begin{array}{ccc}
S_4 & S_1 & S_{14} \\ 
m_4 & m_1 & m_{14}
\end{array}
\right] \left[ 
\begin{array}{ccc}
S_4 & S_1 & S_{14} \\ 
\bar{m}_4 & \bar{m}_1 & \bar{m}_{14}
\end{array}
\right] \delta _{\bar{m}_{23}m_{23}}\delta _{\bar{m}_{14}m_{14}} 
\]
\[
=\sum_{m_1,m_2}\sum_{m_3,m_4}\sum\limits_{m_{12},m_{34}}\left[ 
\begin{array}{ccc}
S_1 & S_2 & S_{12} \\ 
m_1 & m_2 & m_{12}
\end{array}
\right] \left[ 
\begin{array}{ccc}
S_3 & S_4 & S_{34} \\ 
m_3 & m_4 & m_{34}
\end{array}
\right] \left[ 
\begin{array}{ccc}
S_{12} & S_{34} & S \\ 
m_{12} & m_{34} & M
\end{array}
\right] \sum_{\bar{m}_1,\bar{m}_2}\sum\limits_{\bar{m}_3,\bar{m}_4}\left| S_2%
\bar{m}_2\right\rangle \left| S_3\bar{m}_3\right\rangle \left| S_4\bar{m}%
_4\right\rangle \left| S_1\bar{m}_1\right\rangle 
\]
\[
\times \sum\limits_{S_{14},m_{14}}\sum\limits_{S_{23},m_{23}}\sum\limits_{%
\bar{m}_{14},\bar{m}_{23}}\left[ 
\begin{array}{ccc}
S_2 & S_3 & S_{23} \\ 
m_2 & m_3 & m_{23}
\end{array}
\right] \left[ 
\begin{array}{ccc}
S_2 & S_3 & S_{23} \\ 
\bar{m}_2 & \bar{m}_3 & \bar{m}_{23}
\end{array}
\right] \left[ 
\begin{array}{ccc}
S_4 & S_1 & S_{14} \\ 
m_4 & m_1 & m_{14}
\end{array}
\right] \left[ 
\begin{array}{ccc}
S_4 & S_1 & S_{14} \\ 
\bar{m}_4 & \bar{m}_1 & \bar{m}_{14}
\end{array}
\right] 
\]
\[
\times \sum\limits_{\bar{S},\bar{M}}\left[ 
\begin{array}{ccc}
S_{23} & S_{14} & \bar{S} \\ 
m_{23} & m_{14} & \bar{M}
\end{array}
\right] \left[ 
\begin{array}{ccc}
S_{23} & S_{14} & \bar{S} \\ 
\bar{m}_{23} & \bar{m}_{14} & \bar{M}
\end{array}
\right] 
\]

A full contraction of six Clebsch-Gordan coefficients yields the $9j$ symbol 
\[
\sum_{m_1,m_2}\sum_{m_3,m_4}\sum\limits_{m_{12},m_{34}}\left[ 
\begin{array}{ccc}
S_1 & S_2 & S_{12} \\ 
m_1 & m_2 & m_{12}
\end{array}
\right] \left[ 
\begin{array}{ccc}
S_3 & S_4 & S_{34} \\ 
m_3 & m_4 & m_{34}
\end{array}
\right] \left[ 
\begin{array}{ccc}
S_{12} & S_{34} & S \\ 
m_{12} & m_{34} & M
\end{array}
\right] \left[ 
\begin{array}{ccc}
S_4 & S_1 & S_{14} \\ 
m_4 & m_1 & m_{14}
\end{array}
\right] \left[ 
\begin{array}{ccc}
S_2 & S_3 & S_{23} \\ 
m_2 & m_3 & m_{23}
\end{array}
\right] \left[ 
\begin{array}{ccc}
S_{23} & S_{14} & \bar{S} \\ 
m_{23} & m_{14} & \bar{M}
\end{array}
\right] 
\]
\[
=\left( -1\right) ^{S_1+S_4-S_{14}}\left( -1\right) ^{S_3+S_4-S_{34}}\left(
-1\right) ^{S_{23}+S_{14}-\bar{S}}\sum_{m_1,m_2}\sum_{m_3,m_4}\sum%
\limits_{m_{12},m_{34}}\left[ 
\begin{array}{ccc}
S_1 & S_2 & S_{12} \\ 
m_1 & m_2 & m_{12}
\end{array}
\right] \left[ 
\begin{array}{ccc}
S_3 & S_4 & S_{34} \\ 
m_3 & m_4 & m_{34}
\end{array}
\right] \left[ 
\begin{array}{ccc}
S_{12} & S_{34} & S \\ 
m_{12} & m_{34} & M
\end{array}
\right] 
\]
\[
\times \left[ 
\begin{array}{ccc}
S_1 & S_4 & S_{14} \\ 
m_1 & m_4 & m_{14}
\end{array}
\right] \left[ 
\begin{array}{ccc}
S_2 & S_3 & S_{23} \\ 
m_2 & m_3 & m_{23}
\end{array}
\right] \left[ 
\begin{array}{ccc}
S_{14} & S_{23} & \bar{S} \\ 
m_{14} & m_{23} & \bar{M}
\end{array}
\right] 
\]
\[
=\left( -1\right) ^{S_1+S_4-S_{14}}\left( -1\right) ^{S_3+S_4-S_{34}}\left(
-1\right) ^{S_{23}+S_{14}-\bar{S}}\left[ \left( 2S_{12}+1\right) \left(
2S_{14}+1\right) \left( 2S_{23}+1\right) \left( 2S_{34}+1\right) \right]
^{1/2}\left\{ 
\begin{array}{ccc}
S_1 & S_2 & S_{12} \\ 
S_4 & S_3 & S_{34} \\ 
S_{14} & S_{23} & S
\end{array}
\right\} \delta _{S\bar{S}}\delta _{M\bar{M}}, 
\]
where we use the symmetry relation for the Clebsch-Gordan coefficients 
\[
\left[ 
\begin{array}{ccc}
S_1 & S_2 & S_{12} \\ 
m_1 & m_2 & m_{12}
\end{array}
\right] =(-1)^{S_1+S_2-S_{12}}\left[ 
\begin{array}{ccc}
S_2 & S_1 & S_{12} \\ 
m_2 & m_1 & m_{12}
\end{array}
\right] . 
\]
By introducing the state with the recoupled four spins 
\[
\left| S_2S_3(S_{23})S_4S_1(S_{14})\bar{S}\bar{M}\right\rangle =\sum_{\bar{m}%
_1,\bar{m}_2}\sum\limits_{\bar{m}_3,\bar{m}_4}\sum\limits_{\bar{m}_{14},\bar{%
m}_{23}}\left[ 
\begin{array}{ccc}
S_2 & S_3 & S_{23} \\ 
\bar{m}_2 & \bar{m}_3 & \bar{m}_{23}
\end{array}
\right] \left[ 
\begin{array}{ccc}
S_4 & S_1 & S_{14} \\ 
\bar{m}_4 & \bar{m}_1 & \bar{m}_{14}
\end{array}
\right] \left[ 
\begin{array}{ccc}
S_{23} & S_{14} & \bar{S} \\ 
\bar{m}_{23} & \bar{m}_{14} & \bar{M}
\end{array}
\right] 
\]
\[
\times \left| S_2\bar{m}_2\right\rangle \left| S_3\bar{m}_3\right\rangle
\left| S_4\bar{m}_4\right\rangle \left| S_1\bar{m}_1\right\rangle , 
\]
we get finally 
\[
\hat{C}_4\left| S_1S_2(S_{12})S_3S_4(S_{34})SM\right\rangle
=\sum\limits_{S_{14},S_{23}}\left( -1\right) ^{S_1+S_4-S_{14}}\left(
-1\right) ^{S_3+S_4-S_{34}}\left( -1\right) ^{S_{23}+S_{14}-\bar{S}} 
\]
\[
\times \left[ \left( 2S_{12}+1\right) \left( 2S_{14}+1\right) \left(
2S_{23}+1\right) \left( 2S_{34}+1\right) \right] ^{1/2}\left\{ 
\begin{array}{ccc}
S_1 & S_2 & S_{12} \\ 
S_4 & S_3 & S_{34} \\ 
S_{14} & S_{23} & S
\end{array}
\right\} \left| S_2S_3(S_{23})S_4S_1(S_{14})SM\right\rangle , 
\]
that is transformation from a coupling scheme to another. Then an action of
the operator $\hat{C}_4\,$is defined by a linear transformation of the basis 
$\kappa ^{\prime }=\left\{ S_1^{\prime }S_2^{\prime }(S_{12}^{\prime
})S_3^{\prime }S_4^{\prime }(S_{34}^{\prime })\right\} $ 
\[
\hat{C}_4\left| \kappa \right\rangle =\sum\limits_{\kappa ^{\prime
}}D_{\kappa ^{\prime }\kappa }^{(S)}\left( \hat{C}_4\right) \left| \kappa
^{\prime }\right\rangle , 
\]
where the matrix $D_{\kappa ^{\prime }\kappa }^{(S)}\left( \hat{C}_4\right) $
is determined by the expression 
\[
D_{\kappa ^{\prime }\kappa }^{(S)}\left( \hat{C}_4\right) =\delta
_{S_2S_1^{\prime }}\delta _{S_3S_2^{\prime }}\delta _{S_4S_3^{\prime
}}\delta _{S_1S_4^{\prime }}\delta _{S_{23}S_{12}^{\prime }}\delta
_{S_{14}S_{34}^{\prime }}\left[ \left( 2S_{12}+1\right) \left(
2S_{14}+1\right) \left( 2S_{23}+1\right) \left( 2S_{34}+1\right) \right]
^{1/2} 
\]
\[
\times \left( -1\right) ^{S_1+S_4-S_{14}}\left( -1\right)
^{S_3+S_4-S_{34}}\left( -1\right) ^{S_{23}+S_{14}-\bar{S}}\left\{ 
\begin{array}{ccc}
S_1 & S_2 & S_{12} \\ 
S_4 & S_3 & S_{34} \\ 
S_{14} & S_{23} & S
\end{array}
\right\} . 
\]
We can handle analogously another symmetry operations in the $D_4$ group.

\section{Appendix D}

The standard method for constructing an irreducible basis is to use the
projection operator

\begin{equation}
P_\mu ^\Gamma =\frac{\left[ \Gamma \right] }{\left[ G\right] }\sum_{\hat{g}%
\in G}D_{\mu \mu }^{(\Gamma )}(g)\,\hat{g},  \label{Pr1}
\end{equation}
and the shift operator 
\begin{equation}
P_{\mu \nu }^\Gamma =\frac{\left[ \Gamma \right] }{\left[ G\right] }\sum_{%
\hat{g}\in G}D_{\mu \nu }^{(\Gamma )}(g)\,\hat{g},  \label{Pr2}
\end{equation}
where $\left[ G\right] $ is the order of the group $G$, $\left[ \Gamma
\right] $ is the dimension of the irreducible representation $\Gamma $ and $%
D_{\mu \nu }^{(\Gamma )}(g)$ are the irreducible matrix elements, $\mu $ or $%
\nu $ is an index enumerating the basis. Supposing that $\psi $ is one of
the reducible basis vectors of $G$, an irreducible basis might be obtained
by applying 
\begin{equation}
P_\mu ^\Gamma \psi =\left( \psi _\mu ^\Gamma \cdot \psi \right) \psi _\mu
^\Gamma .  \label{Act1}
\end{equation}
If $\left\{ \psi _\nu ^\Gamma \right\} $ is the basis for the irrerp $\Gamma 
$ then 
\begin{equation}
P_{\mu \nu }^\Gamma \psi _\nu ^\Gamma =\psi _\mu ^\Gamma .  \label{Act2}
\end{equation}
Let us construct irreducible tensors $U_{q\mu }^{1\Gamma }$ from the
operators $\left\{ S_a,S_b,S_c,S_d\right\} $ forming the nearest environment
of the central site. The transformations of one of the given spins under the
elements of the group $D_4$ are 
\[
E\,S_a=S_a,\;C_4\,S_a=S_d,\;C_4^2\,S_a=S_c,\;C_4^3\,S_a=S_b, 
\]
\[
C_2^x\,S_a=S_b,\;C_2^y\,S_a=S_d,\;C_v^{^{\prime }}\,S_a=S_a,\;C_v^{^{\prime
\prime }}\,S_a=S_c, 
\]
that together with (\ref{Pr1}) gives immediately the irreducible tensors $%
U_{q\mu }^{1\Gamma }$ of the one-dimensional representations 
\[
U_{q1}^{1A_1}=N_{A_1}\left( S_a+S_b+S_c+S_d\right) ,\;U_{q1}^{1A_2}=0, 
\]
\[
U_{q1}^{1B_1}=0,\;U_{q1}^{1B_2}=N_{B_2}\left( S_a-S_b+S_c-S_d\right) . 
\]
To find the irredicible basis for the two-dimensional representation $E$ we
construct the projection operator $P_1^E$ and then apply it to the spin $S_a$
that yields 
\[
U_{q1}^{1E}=N_E\left( S_a+S_b-S_c-S_d\right) . 
\]
Using the shift operator (\ref{Pr2}) and acting according to the rule (\ref
{Act2}) we obtain the second irrep basis vector 
\[
U_{q2}^{1E}=N_E\left( S_a-S_b-S_c+S_d\right) . 
\]
We choose the coefficents $N_\Gamma $ ($\Gamma =A_1$, $E$) so that
Hamiltonian written through the irreducible tensors coincides with the
initial spin operator form.

It is not always possible to construct all irredicible basises from one
chain. The theory says that one have to choose another starting function. In
a computer realization, therefore, we build all chains generated by all
vectors of a reducible basis 
\begin{equation}
\hat{g}\psi _i=\sum_jD_{ji}\left( \hat{g}\right) \psi _j,  \label{chain}
\end{equation}
and form the matrix $\hat{X}^\Gamma $ from the chains 
\begin{equation}
\hat{X}_{ji}^\Gamma =N_\Gamma \sum_gD_{\mu \mu }^\Gamma \left( \hat{g}%
\right) D_{ji}\left( \hat{g}\right) .  \label{Xop}
\end{equation}
The rank of this matrix 
\[
C_\Gamma =\frac 1{\left[ \Gamma \right] }\sum_g\chi \left( g\right) \chi
^\Gamma \left( g\right) 
\]
determines a number of linear independent columns, where the character $\chi
\left( \hat{g}\right) =\sum_iD_{ii}\left( \hat{g}\right) $. After these
columns are established, orthogonalized and normalized with the help of
Schmidt-Gram procedure we get the first $C_\Gamma $ columns of the
transformation matrix $\hat{T}_{i,\Gamma \mu }$. By running over all
irreducible representations and repeating the basic steps in the approach we
obtain the square matrix of corresponding similarity transformation onto the
symmetry adapted basis 
\[
\psi _\mu ^\Gamma =\sum_i\hat{T}_{i,\Gamma \mu }\psi _i. 
\]
As an example we calculate $\hat{T}_{S_{12}S_{34},\Gamma \mu }^{(1)}$ for
the nearest-neighbor environment of the central site. The characters of
3-dimensional representation can be read off from the $3\times 3$ matrices $%
D^{(1)}$ given in Sec. III We thus obtain 
\[
\begin{array}{cccccc}
& E & C_4,C_4^3 & C_4^2 & C_2^x,C_2^y & \sigma _v^{^{\prime }},\sigma
_v^{^{\prime \prime }} \\ 
\chi & 3 & -1 & -1 & -1 & 1
\end{array}
\]
whence we conclude $D^{(1)}=D^{(1B_1)}\oplus D^{(1E)}$. A direct calculation
of $\hat{X}^\Gamma $ matrices from Eq.(\ref{Xop}) yields

\[
\hat{X}_{ji}^{B_1}=\frac 18\sum_g\chi ^{B_1}\left( \hat{g}\right)
D_{ji}^{(1)}\left( \hat{g}\right) =\left( 
\begin{array}{ccc}
\frac 12 & \frac 12 & 0 \\ 
\frac 12 & \frac 12 & 0 \\ 
0 & 0 & 0
\end{array}
\right) , 
\]
\[
\hat{X}_{11}^E=\frac 28\sum_gD_{11}^E\left( \hat{g}\right)
D_{ji}^{(1)}\left( \hat{g}\right) =\left( 
\begin{array}{ccc}
\frac 14 & -\frac 14 & \frac{\sqrt{2}}4 \\ 
-\frac 14 & \frac 14 & -\frac{\sqrt{2}}4 \\ 
\frac{\sqrt{2}}4 & -\frac{\sqrt{2}}4 & \frac 12
\end{array}
\right) , 
\]
and 
\[
\hat{X}_{21}^E=\frac 28\sum_gD_{21}^E\left( \hat{g}\right)
D_{ji}^{(1)}\left( \hat{g}\right) =\left( 
\begin{array}{ccc}
\frac 14 & -\frac 14 & \frac{\sqrt{2}}4 \\ 
-\frac 14 & \frac 14 & -\frac{\sqrt{2}}4 \\ 
-\frac{\sqrt{2}}4 & \frac{\sqrt{2}}4 & -\frac 12
\end{array}
\right) . 
\]
By noticing that ranks of the matrices equal to unity, we find finally via
the Schmidt-Gram procedure the transformation matrix $\hat{T}_{S_{\alpha
\beta }S_{\gamma \eta };\Gamma \mu }^{(1)}$ 
\[
\begin{array}{cccc}
& \left| 1M;B_1\right\rangle & \left| 1M;E1\right\rangle & \left|
1M;E2\right\rangle \\ 
\left| 01;1M\right\rangle & \frac 1{\sqrt{2}} & \frac 12 & \frac 12 \\ 
\left| 10;1M\right\rangle & \frac 1{\sqrt{2}} & -\frac 12 & -\frac 12 \\ 
\left| 11;1M\right\rangle & 0 & \frac 1{\sqrt{2}} & -\frac 1{\sqrt{2}}
\end{array}
\text{.} 
\]

\section{Appendix E}

For the reader convenience we give the character table of the group $D_4$%
\[
\begin{array}{cccccc}
D_4 & E & C_4,C_4^{-1} & C_4^2 & C_2^x,C_2^y & C_v^{^{\prime
}},C_v^{^{\prime \prime }} \\ 
A_1 & 1 & 1 & 1 & 1 & 1 \\ 
A_2 & 1 & 1 & 1 & -1 & -1 \\ 
B_1 & 1 & -1 & 1 & 1 & -1 \\ 
B_2 & 1 & -1 & 1 & -1 & 1 \\ 
E & 2 & 0 & -2 & 0 & 0
\end{array}
\]

and the matrices of double irreducible representation taken in the basis $xy$
(see \cite{Koster}, for example)

$D^{(E)}(E)=\left( 
\begin{array}{cc}
1 & 0 \\ 
0 & 1
\end{array}
\right) ,$ $D^{(E)}(C_4)=\left( 
\begin{array}{cc}
0 & -1 \\ 
1 & 0
\end{array}
\right) ,$ $D^{(E)}(C_4^2)=\left( 
\begin{array}{cc}
-1 & 0 \\ 
0 & -1
\end{array}
\right) ,$ $D^{(E)}(C_4^3)=\left( 
\begin{array}{cc}
0 & 1 \\ 
-1 & 0
\end{array}
\right) ,$

$D^{(E)}(C_2^x)=\left( 
\begin{array}{cc}
1 & 0 \\ 
0 & -1
\end{array}
\right) ,$ $D^{(E)}(C_2^y)=\left( 
\begin{array}{cc}
-1 & 0 \\ 
0 & 1
\end{array}
\right) ,$ $D^{(E)}(C_v^{^{\prime }})=\left( 
\begin{array}{cc}
0 & 1 \\ 
1 & 0
\end{array}
\right) ,$ $D^{(E)}(C_v^{^{\prime \prime }})=\left( 
\begin{array}{cc}
0 & -1 \\ 
-1 & 0
\end{array}
\right) $.

\newpage

Fig.1 Clusters used in the calculations.

Fig.2 The lowest-energy spectrum of the environment for the cluster $\sqrt{17%
}\times \sqrt{17}$ on the square lattice. The SU(2) symmetry breaks and a
long-range N\'{e}el order appears as a set of $A_1$-states with an energy
scaling as $E(S)\sim S(S+1)$ (dashed line). The symbols represent the
irreducible representations of the different eigenstates.

Fig.3 The cluster ground state energy $E$, the energy per bond $\varepsilon $%
, and the staggered magnetization $m$ convergence for the $\sqrt{17}\times 
\sqrt{17}$ cluster vs number of environment states kept.


\begin{references}
\bibitem{Dagotto}  E. Dagotto, Rev. Mod. Phys. {\bf 66}, 763 (1994).

\bibitem{White}  S.R. White, Phys. Rev. Lett. {\bf 69}, 2863 (1992); Phys.
Rev. B {\bf 48}, 10345 (1993);

\bibitem{Peschel}  I. Peschel, K. Hallberg, X. Wang, and M. Kaulke, 1999,
Eds., {\it Density Matrix Renormalization: a New Numerical Method}, Lecture
Notes in Physics No. 528 (Springer, New York).

\bibitem{Schollwock}  U. Schollw\"{o}ck, Rev. Mod. Phys.{\bf 77}, 259 (2005).

\bibitem{Xiang}  T. Xiang, J.Z. Lou, and Z.B. Su, Phys. Rev. B {\bf 64},
104414 (2001).

\bibitem{Farnell}  D.J.J. Farnell, Phys. Rev. B {\bf 68}, 134419 (2003).

\bibitem{Sandvik}  A.W. Sandvik, Phys. Rev. B {\bf 56}, 11678 (1997).

\bibitem{Nishino}  G. Sierra and T. Nishino, Nucl. Phys. B{\bf \ 495}, 505
(1997).

\bibitem{Wada}  W. Tatsuaki, Phys. Rev. E {\bf 61}, 3199 (2000).

\bibitem{McCulloch}  I.P. McCulloch, M. Gulasci, Aust. J. Phys. {\bf 53},
597 (2000).

\bibitem{Gulasci}  I.P. McCulloch, M. Gulasci, Europhys. Lett. {\bf 57}, 852
(2002).

\bibitem{Ostlund}  S. Ostlund and S. Rommer, Phys. Rev. Lett. {\bf 75}, 3537
(1995); Phys. Rev. B {\bf 55}, 2164 (1997).

\bibitem{Dukelsky}  J. Dukelsky, M.A. Mart\'{\i}n-Delgado, T. Nishino, and
G. Sierra, Europhys. Lett. {\bf 43}, 457 (1998).

\bibitem{Roman}  J.M. Roman, G. Sierra, J. Dukelsky, and M.A.
Mart\'{\i}n-Delgado, J. Phys. A {\bf 31}, 9729 (1998).

\bibitem{Manousakis}  E. Manousakis, Rev. Mod. Phys. {\bf 63}, 1 (1991).

\bibitem{Barnes}  T. Barnes, J. Mod. Phys. C {\bf 2}, 659 (1991).

\bibitem{Lin}  H.-Q. Lin, J.S. Flynn, D.D. Betts, Phys. Rev. B {\bf 64},
214411 (2001).

\bibitem{Campbell}  H.Q. Lin, D.K. Campbell, Phys. Rev. Lett. {\bf 69}, 2415
(1992).

\bibitem{Malrieu}  J.P. Malrieu and N. Guih\'{e}ry, Phys. Rev. B {\bf 63},
085110 (2001).

\bibitem{Wind}  P. Wind, N. Guih\'{e}ry, and J.P. Malrieu, Phys. Rev. B {\bf %
59}, 2556 (1999).

\bibitem{Hajj}  M.A. Hajj, N. Guih\'{e}ry, and J.P. Malrieu, P. Wind, Phys.
Rev. B {\bf 70}, 094415 (2004).

\bibitem{Bishop}  R.F. Bishop, J.B. Parkinson, and Y. Xian,Phys.Rev. B {\bf %
43}, 13782 (1991).

\bibitem{Hale}  R.F. Bishop, R.G. Hale, and Y. Xian, Phys. Rev. Lett. {\bf 73%
}, 3157 (1994).

\bibitem{Cheng}  H.Q. Lin, D.K. Campbell, Y.C. Cheng, and C.Y. Pan, Phys.
Rev. B {\bf 50}, 12701 (1994).

\bibitem{Masui}  D.D. Betts, S. Masui, N. Vats, and G.E. Stewart, Can. J.
Phys. {\bf 74}, 54 (1996).

\bibitem{Zeng}  C. Zeng, D.J.J. Farnell,and R.F. Bishop, J. Stat. Phys. {\bf %
90}, 327 (1998).

\bibitem{Betts}  D.D. Betts, H.Q. Lin, J.S. Flyn, Can. J. Phys.{\bf \ 77},
353 (1999).

\bibitem{Hasenfratz}  P. Hasenfratz and F. Niedermayer, Z. Phys. B: Condens.
Matter {\bf 92}, 91~(1993).

\bibitem{Misguich}  G. Misguich, C. Lhuillier, and B. Bernu, Phys. Rev. B 
{\bf 60}, 1064 (1999).

\bibitem{Haan}  O. Haan, J.-U. Klaetke, and K.-H. M\"{u}tter, Phys. Rev. B 
{\bf 46}, 5723 (1992).

\bibitem{Mattis}  E.H. Lieb and D.C. Mattis, J. Math. Phys. {\bf 3}, 749
(1962).

\bibitem{Koster}  G.F. Koster, J.O. Dimmock, R.G. Wheeler and H. Statz, {\it %
Properties of the Thirty Two Point Groups}, (M.I.T. Press, Cambridge, 1963).

\bibitem{Varshalovich}  D.A. Varshalovich, A.N. Moskalev, V.K. Khersonskii, 
{\it Quantum theory of angular momentum} (World Scientific, 1988).

\bibitem{Zittartz}  H. Niggemann, A. Kl\"{u}mper and J. Zittartz, Z. Phys. B 
{\bf 104}, 103 (1997); M.A. Ahrens, A. Schadschneider, J. Zittartz, Phys.
Rev. B {\bf 71}, 174432 (2005).

\bibitem{Nishino1}  Nishino, T., Y. Hieida, K. Okunishi, N. Maeshima, Y.
Akutsu, and A. Gendiar, Prog. Theor. Phys. {\bf 105}, 409 (2001); the spin-$%
1/2$ AFH\ model on a square lattice is considered in Y. Nishino, N.
Maeshima, A. Gendiar, and T. Nishino, cond-mat/0401115.

\bibitem{Martin}  M.A. Mart\'{\i}n-Delgado, M. Roncaglia, and G. Sierra,
Phys. Rev. B {\bf 64}, 075117 (2001).

\bibitem{Elliot}  J.P. Elliot, P.G. Dawber, {\it Symmetry in Physics}
(Macmillan, London, 1979).

\bibitem{Griffith}  G.S. Griffith, {\it The irreducible Tensor Method for
Molecular Symmetry Groups}, (New Jersey, 1962).
\end{references}
\end{document}